# Sample volume as a key design parameter in affinity-based biosensors

Daan Beijersbergen[1,2], Jérôme Charmet[1,2]

Correspondence to: jerome.charmet@he-arc.ch

1. School of Engineering–HE-Arc Ingénierie, HES-SO University of Applied Sciences Western Switzerland, Neuchâtel, Switzerland
2. School of Precision and Biomedical Engineering, University of Bern, Bern, Switzerland

## Abstract

Affinity-based biosensors have become indispensable in modern diagnostics and health monitoring. While considerable research has focused on optimizing analyte transport and binding kinetics, a fundamental parameter—sample volume—remains largely underexplored in biosensor design. This is critical because biosensor performance depends on the absolute number of target molecules present, not solely their concentration, making volume a key consideration where sample availability is limited.

To address this gap, we developed a mathematical two-compartment model integrating simplified mass transport, Langmuir binding kinetics, and mass conservation under finite volume constraints. The model accurately simulates biosensor binding kinetics (root-mean-square deviation < 5%) and predicts equilibration time and required volume (mean relative error of 7.3%) compared to finite-element simulations, whilst achieving more than 100-fold reduction in computational time.

From the framework, we derived analytical expressions for biosensor equilibration time and required volume as a function of the Damköhler number, ranging from reaction-limited to transport-limited systems. These analytical solutions predict equilibration time and required volume for a biosensors (mean relative error of 11.0% and 11.7% compared to numerical simulations), providing rapid estimates from first-order biosensor parameters without numerical simulation.

We validated this framework experimentally by optimizing flow rate parameters for a quartz crystal microbalance (QCM) biosensor and retrospectively applied optimization guidelines on a published biosensor. The open-source model and analytical expressions allow researchers to gain mechanistic insights, optimize device performance, and make informed design decisions tailored to specific healthcare contexts, including point-of-care testing and resource-constrained environments.

# Introduction

Affinity-based biosensors are revolutionizing clinical diagnostics by enabling rapid, cost-effective detection of target molecules in biofluids—capabilities that are transforming healthcare delivery from centralized laboratories to point-of-care settings [1,2]. These devices have already proven their clinical value in diverse applications, for example in measuring prostate-specific antigen (PSA) for prostate cancer screening or detecting pTau for early Alzheimer's disease diagnosis in clinically relevant volumes [3–5]. With their fast readout times, operational simplicity, and dramatically lower costs compared to traditional methods such as polymerase chain reaction (PCR), biosensors are a powerful tool for point-of-care applications in resource-limited settings, emergency departments, and home monitoring scenarios where timely decisions can be life-saving 7,8].

In affinity-based biosensors, target molecules are transported to the sensor's surface where they bind to complementary bioreceptors, such as antibodies, enzymes, or aptamers, inducing signals through a transducer [9]. While transport optimization has been recognized as critical to biosensor performance [10–12], a striking gap persists in how biosensors are conceptualized and designed. Current depictions typically include only the target molecule, bioreceptors (bio-interface), transducers, and signal processing components [13–17], treating transport as an afterthought rather than a fundamental design element. This oversight limits our ability to rationally optimize biosensors for real-world clinical constraints. To address this critical gap, we propose integrating transport explicitly as a core architectural component of affinity-based biosensors (Figure 1a).

Within this transport-integrated framework, sample volume emerges as a parameter with profound implications for biosensor design and clinical utility. The fundamental constraint is straightforward yet frequently overlooked: the total number of target molecules available for detection, $N = cV$, depends equally on concentration, $c$ and volume, $V$. This has immediate practical consequences. Sample volumes vary dramatically by bodily source—from microliters for sweat or lacrimal fluid to milliliters for venous blood or urine [18]. Clinical context further constrains available volume: point-of-care finger-prick sampling yields only microliters, while venous collection in clinical laboratories can provide milliliters. Despite these well-known limitations, concentration is routinely reported without volume in the literature [3], and volume considerations are absent even in studies explicitly focused on transport optimization [11,12]. This gap between biosensor theory and clinical reality undermines our ability to design devices fit for their intended healthcare setting.

The consequences of neglecting volume constraints become apparent when considering optimization strategies. For large-volume, high-concentration samples with abundant target molecules, transport and binding optimization may be less critical: devices can be operated sub-optimally without compromising performance. However, volume-limited scenarios present fundamentally different challenges. While increasing flow rate effectively reduces measurement time for large-volume samples, this strategy can fail catastrophically for small volumes where sample depletion occurs before measurement completion. This raises a question with direct clinical impact: when and how should biosensors be optimized given practical volume constraints that vary across healthcare settings?

To answer this question and provide actionable design guidance, we developed a two-compartment biosensor model that incorporates simplified mass transport and classical Langmuir binding while enforcing mass conservation under finite-volume conditions. The model provides mechanistic insight into biosensing performance and was validated against numerical simulations. From this open-source model, we derived analytical expressions for the equilibration time and required volume, which provide rapid insights into biosensor performance without numerical simulation.

This paper is roughly divided into three parts. First, we introduce individual aspects of affinity-based biosensor, including binding and transport. Second, we discuss our implementation of the two-compartment model and showcase the characterization of an experimental biosensor. Third, we discuss model implications, including analytical expressions and biosensor optimization. In total, this paper offers researchers and engineers versatile tools for understanding and optimizing biosensor kinetics across the full spectrum of healthcare applications, from well-resourced clinical laboratories to point-of-care settings in low-resource environments.

# Main text

## Biosensors rely on an interplay between transport and binding

In affinity-based biosensors, target molecules must reach the transducer surface and bind to bioreceptors to generate a measurable signal (Figure 1a). However, rational biosensor optimization remains a significant challenge: analyte transport and receptor–ligand binding are governed by a large set of coupled parameters, making systematic design inherently complex and often reliant on trial-and-error approaches that delay clinical translation.

Accurate modeling of biosensor performance typically requires computationally intensive numerical simulations to account for local concentration gradients, depletion effects and time-varying flow parameters. This creates a critical barrier to widespread biosensor optimization: commercial finite-element packages demand substantial computational resources and expertise, while even open-source alternatives [19,20] require specialized knowledge and remain prohibitively time-consuming for routine design iteration. This computational bottleneck limits the ability of researchers, particularly in resource-constrained settings, to rapidly prototype and optimize biosensors for specific clinical applications.

We address this challenge by leveraging a key physical insight: biosensors generally operate at low Reynolds numbers due to small characteristic length scales, leading to laminar flow regimes that enable dramatic model simplification. This allows us to develop a tractable analytical framework with only thirteen governing parameters (Figure 1c), each with clear physical meaning and direct experimental control (Figure 1b). This reduction transforms biosensor optimization from a computationally restrictive problem to one amenable to rapid exploration and mechanistic understanding.

The critical performance metric in affinity-based biosensors is target molecule binding at the sensor surface, an outcome determined by the interplay between mass transport to the interface and subsequent binding kinetics. To develop predictive design rules, we must first establish how these two fundamental processes operate independently before examining their coupled behavior under volume-limited conditions.

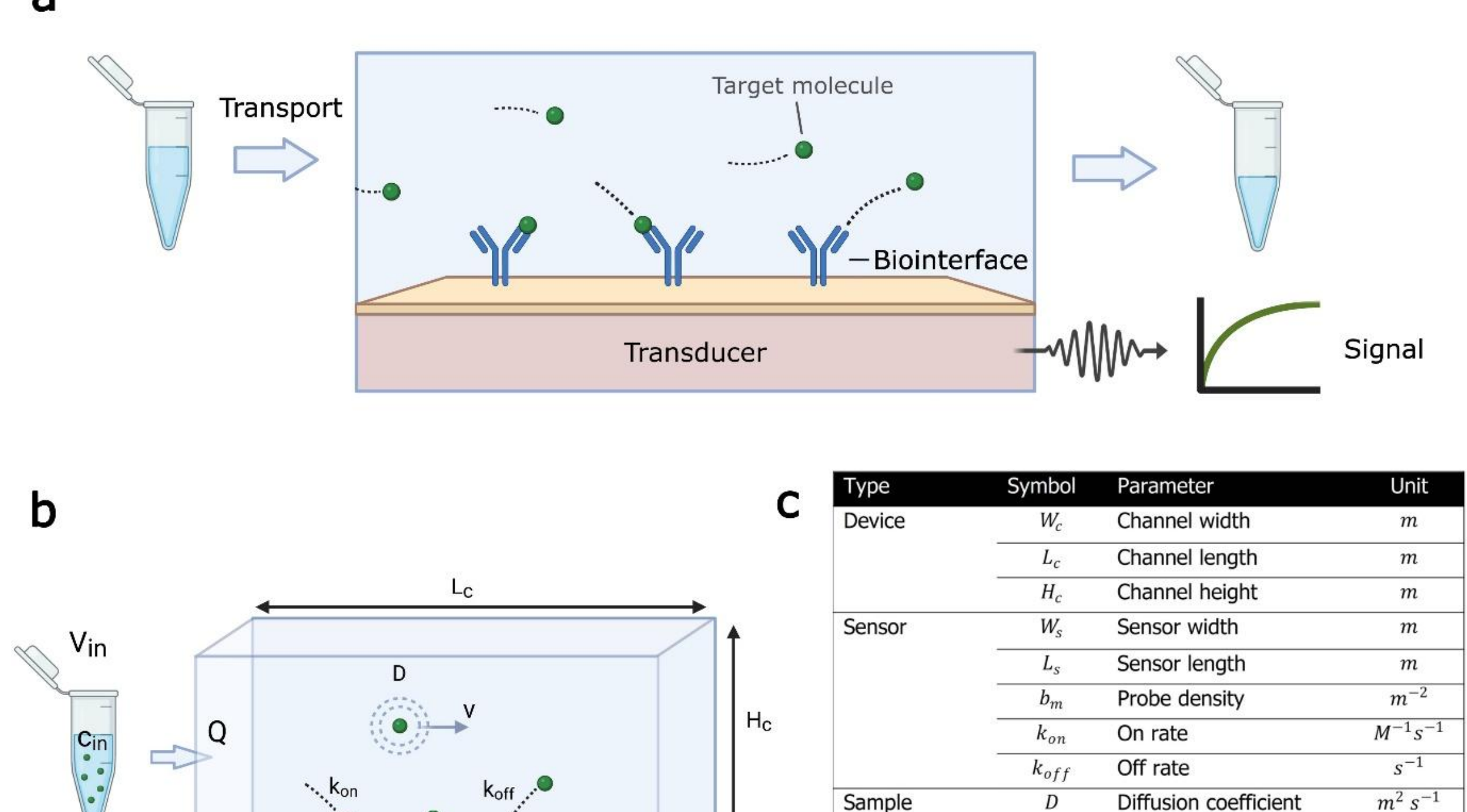

| Type | Symbol | Parameter | Unit |
|---|---|---|---|
| Device | $W_c$ | Channel width | $m$ |
| | $L_c$ | Channel length | $m$ |
| | $H_c$ | Channel height | $m$ |
| Sensor | $W_s$ | Sensor width | $m$ |
| | $L_s$ | Sensor length | $m$ |
| | $b_m$ | Probe density | $m^{-2}$ |
| | $k_{on}$ | On rate | $M^{-1}s^{-1}$ |
| | $k_{off}$ | Off rate | $s^{-1}$ |
| Sample | $D$ | Diffusion coefficient | $m^2\ s^{-1}$ |
| | $c_{in}$ | Input concentration | $M$ |
| | $c_0$ | Starting concentration | $M$ |
| | $V_{in}$ | Input volume | $m^3$ |
| | $Q$ | Flow rate | $m^3\ s^{-1}$ |

***Figure 1*** *Contextual biosensor definition and overview. (**a**) General biosensor overview. Volume $V_{in}$ contains target molecules that are transported past the biosensor and that can, in the right conditions, bind to bioreceptors. The transducer element converts binding events into a signal. For biosensors, each crucial element needs to be taken into account: transport, biointerface, transducer and signal processing. (**b, c**) A set of 13 parameters are used to describe our biosensor model, which considers the transport and binding elements collectively.*

## Binding rate in a biosensor

Target molecules that are transported close to the surface can reversibly bind to the bioreceptor. This volume close to the surface is considered the 'interface'. The binding and unbinding kinetics are described by the classical first-order Langmuir model, which relates surface coverage to association and dissociation rate constants [11,21]:

$$\frac{db}{dt} = k_{on}c_s(b_m - b) - k_{off}b \quad (1)$$

With $c_s$: the interface concentration $(mol/L)$, $b$: the density of bound target molecules $(mol/m^2)$, $k_{on}$ the on-rate and $k_{off}$ the off-rate. The binding affinity is given by:

$$k_D = \frac{k_{off}}{k_{on}} \quad (2)$$

The binding rate is maximal initially when the surface is unoccupied $(b = 0)$ and decreases as surface sites become saturated until equilibrium is reached $(db/dt = 0)$. As the interface concentration approaches the input concentration $(c_s = c_{in})$, the bound surface equilibrium density $b_{eq}$ is :

$$b_{eq} = \frac{k_{on}c_{in}b_m}{k_{on}c_{in}+k_{off}} \quad (3)$$

Equilibrium binding thus depends on sensor characteristics $(k_{on}, k_{off}, b_m)$ and input concentration $(c_{in})$. This concentration dependence underlies the working principle of affinity-based biosensors: higher input concentrations yield greater surface binding and stronger signals. The nonlinear relationship in Equation 3 necessitates calibration curves to correlate concentration with surface binding. Consequently, reaching equilibrium is essential for accurate concentration determination.

For zero initial surface coverage $c_0 = 0$, Equation 1 yields:

$$b(t) = b_{eq}(1 - exp.(-(k_{on}c_{in} + k_{off})t) \quad (4)$$

Measurement time is defined as the time to reach equilibrium binding, $b_{eq}$. We define equilibration time as $\mathrm{t_{eq}}$when $\mathrm{b(t_{eq}) = 0.95b_{eq}}$. Under conditions where binding does not deplete the interface concentration (analyte abundance), equilibration time is:

$$t_{eq} = \frac{-\ln(0.05)}{k_{on}c_{in}+k_{off}} \quad (5)$$

Where the binding rate $t_R = k_{on}c_{in} + k_{off}\ (s^{-1})$ governs equilibration kinetics. Notably, equilibration time is independent of maximum surface density $b_m$.

## Sufficient volume to reach equilibrium

While concentration can theoretically be inferred before equilibrium, this approach is considerably more complex [22]. Ideally, biosensors should ensure sufficient input molecules $N_{in} = c_{in}V_{in}$to reach $b_{eq}$.

The minimum number of molecules required for equilibrium is $N_{req} = b_{eq}S$, where $S = W_sL_s$ is the sensor surface area. This defines the minimum required volume:

$$V_{min} = \frac{N_{req}}{c_{in}} = \frac{b_{eq}S}{c_{in}} = \frac{W_sL_sk_{on}b_m}{k_{on}c_{in}+k_{off}} \ (6)$$

In practice, larger volumes are needed to account for concentration depletion and dissociation kinetics.

This seemingly simple requirement is crucial for biosensor design. Biomedical sample volumes vary dramatically by source and clinical setting: from microliters for finger-prick or tear samples to milliliters for urine or venous blood. Target concentrations span an equally broad range ($pM$ to $\mu M$), creating diverse operational scenarios (Figure 2a). Biosensor design must therefore be tailored to specific concentration-volume combinations, recognizing that detection depends on absolute molecule number $N = cV$, not concentration alone.

Figure 2b illustrates this volume dependence experimentally. A 100 nM EpCAM (target molecule) solution measured by quartz crystal microbalance (QCM) reaches equilibrium with $200\,\mu L$ but not with $50\,\mu L$. Another important point can be made here. When considering volume only, Equation 6 predicts a minimum required volume of $43\,\mu L$. Indeed, the $50\,\mu L$ sample contains $5.0 \cdot 10^{-12}$ moles, exceeding the theoretical requirement of $4.3 \cdot 10^{-12}$ moles, yet fails to reach equilibrium. This discrepancy highlights that transport kinetics and reversible binding must be considered alongside total molecule availability. Our two-compartment model (dotted lines, Figure 2b) accurately captures this interplay, demonstrating that $V_{min}$ represents a lower bound that must be adjusted for real-world transport limitations and binding-unbinding dynamics.

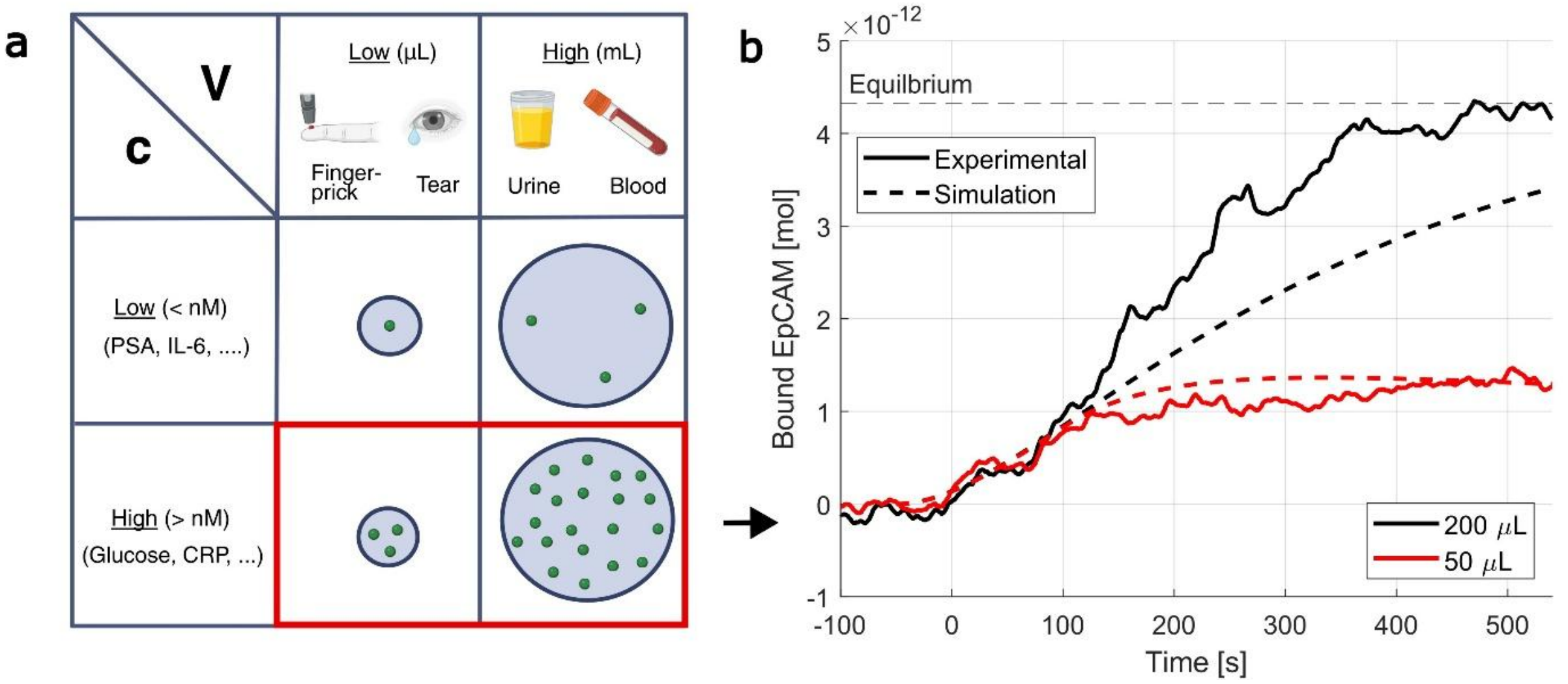

***Figure 2*** *(**a**) Schematic of various clinical sample situations with biomarkers indicated in green. Samples are subdivided in small and large volume, and low and high concentrations. (**b**) Experimental QCM and simulated binding kinetics for 100 nM EpCAM protein binding to aptamers after injection at t = 0. For an identical flow rate of 20* $\mu$L/min, 50 $\mu$L *is insufficient volume to reach the equilibrium for this biosensor.*

## Simplified transport to the interface

Target molecule transport occurs through diffusion and convection. Diffusion follows a random-walk process with characteristic timescale $t_{diff} \sim L^2/D$, where $D$ is the diffusion coefficient. Convection is flow-induced with timescale $t_{conv} \sim L/u$, where $u$ is the velocity and $L$ is the characteristic length scale. The average velocity parallel to the sensor is:

$$u_L = \frac{Q}{W_c H_c} \quad (7)$$

The Péclet number quantifies the relative importance of convection versus diffusion. Using channel height as the characteristic length ($L = H_c$):

$$Pe_H = \frac{t_{diff}}{t_{conv}} = \frac{H_c^2/D}{H_c/u_L} = \frac{Q}{W_c D} \quad (8)$$

Convection dominates for $Pe_H \gg 1$, diffusion dominates for $Pe_H \ll 1$. In diffusion-dominated systems, all molecules reach the interface before passing the sensor surface, but transport is slow. Flow significantly accelerates transport, but excessive flow (convective timescale < diffusive timescale) could cause molecule loss. In laminar flow, transport orthogonal to the sensor surface occurs solely by diffusion. For target molecules to bind to the surface, they must diffuse downward before being swept away. This defines a critical flow rate $Q_c$, where $Q \leq Q_c$ ensures complete delivery: all molecules can potentially bind, subject to Langmuir kinetics. The value of $Q_c$ depends on specific biosensor parameters, thus showcasing the complexity of modeling target molecule transport. A perspective paper by Squires et al. [11] provide a solution by simplifying transport using dimensionless parameters. In particular, they introduced a second Péclet number $Pe_s$:

$$Pe_s = 6\lambda^2 Pe_H \quad (9)$$

With $\lambda = L_s/H_c$ the ratio of sensor length to channel height. These Péclet numbers $Pe_H, Pe_s$ are used to relate to the dimensionless flux to the surface $F$ (the Sherwood number). From literature, the limiting cases are:

$F \approx Pe_H$ for $Pe_H \ll 1$

$F \approx 0.81 Pe_s^{1/3} + 0.71 Pe_s^{-1/6} - 0.2 Pe_s^{-1/3} \ldots$ for $Pe_s \gg 1$ [23]

$F \approx \pi \left( \ln \left( 4/Pe_s^{1/2} \right) + 1.06 \right)^{-1}$ for $Pe_H \gg 1$ and $Pe_s \ll 1$ [24]

Most biosensors operate between these extremes. A continuous approximation via interpolation is detailed in **Supplementary B**. This leads to an approximation of dimensionless flux given the system characteristics of $Pe_h$ and $\lambda$ as seen in Figure 3a. The color bar indicates $F/Pe_h$ ratio, marking the complete delivery regime. The white star corresponds to conditions in Case study 2, representing a biosensor outside the complete delivery regime.

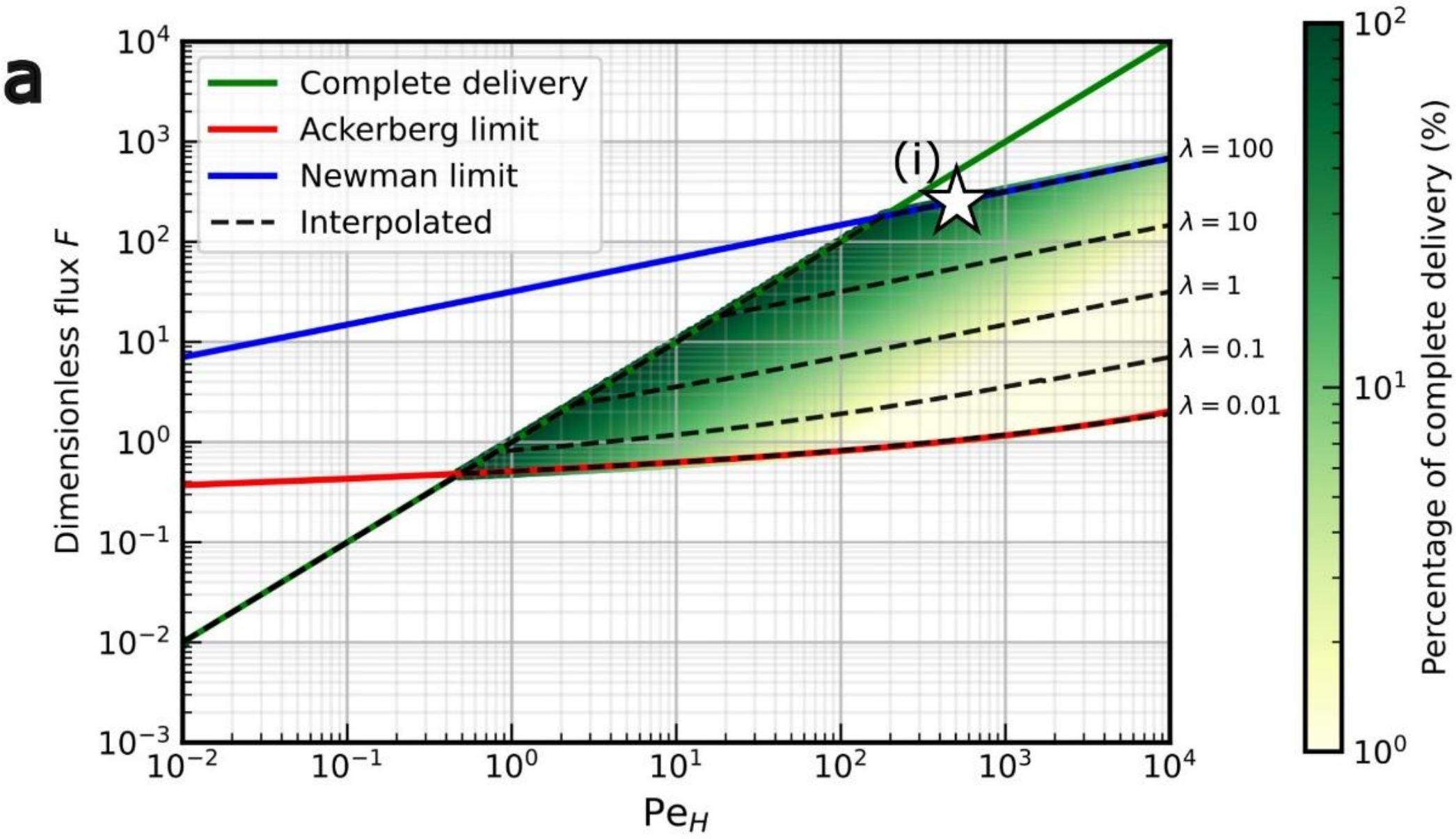

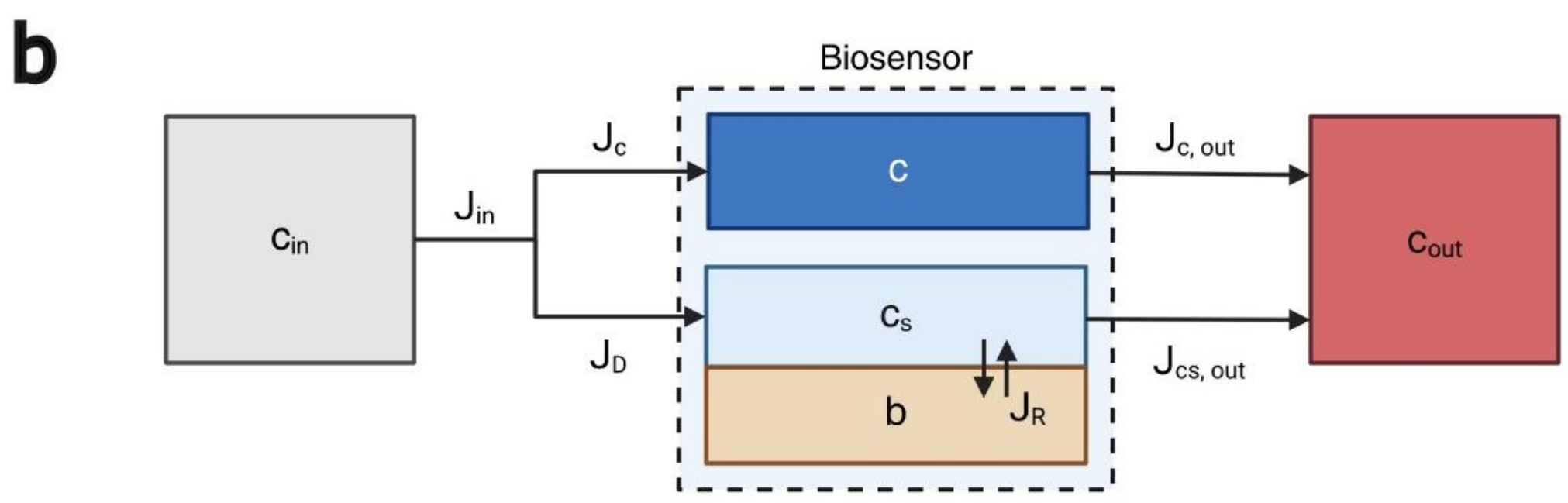

***Figure 3*** *(**a**) Dimensionless flux for mass transport to the interface [11]. For $Pe_H$ << 1, there is complete delivery of molecules onto the biosensor. For $Pe_H$ > 1, the flux depends on the ratio $\lambda$ ($L_s/H_c$), where a higher ratio lambda performs better. The lower extremity is defined by $\lambda$ = 0.01 (from Ackerberg et al. [23]) and the upper extremity is defined by $\lambda$ = 100 (from Newman [24]). The dimensionless flux is interpolated from the extremities for the values in between. The percentage of complete delivery shows what part of the input molecules arrive at the interface. The star (i) corresponds with the case study of Case study 2 and shows a deviation from the complete delivery regime. (**b**) Two-compartment flow chart. The input sample contains concentration $c_{in}$ and flows into the interface $c_s$ or the non-interacting bulk $c$. Target molecules in $c_s$ are able to bind to the surface $b$.*

The Sherwood number yields the dimensional transport rate $k_m (m/s)$ to the biosensor interface [25]

$$k_m = \frac{DF}{H_c} \quad (10)$$

This single parameter simplifies transport to the interface, with flux $J_D = k_m c_{in} W_s H_c$ (mol/s).

## The two-compartment model simulates transport and binding

With an understanding of the transport and binding of biosensors, we can now implement these aspects in a two-compartment model to simulate biosensors [26] (Figure 3b). The biosensor volume is divided into an interacting interface (concentration $c_s$) and a non-interacting bulk (concentration $c_b$), both assumed well-mixed. Target molecules enter with flux $J_{in} = c_{in}Q\ (mol/s)$. Interfacial flux is $J_D = k_m c_{in} W_s H_c\ (mol/s)$. Under 'complete delivery' ($F = Pe_h$), all molecules enter the interface ($J_D = J_{in}$). Beyond this limit ($F < Pe_h$), $J_D < J_{in}$ and target molecules also populate the bulk compartment, following conservation of mass: $J_b = J_{in} - J_D$.

Interface concentration $c_s$ drives binding flux $J_R = \frac{db}{dt} W_s L_s$ via Langmuir kinetics. Unbound molecules exit the compartments with respective fluxes $J_{cs,out} = k_m c_s W_s H_c$ and $J_{cb,out} = (Q - k_m W_s H_c) c_b$.

For a sensor spanning the width of the channel, the respective volume of the two compartments are given by $V_s = \delta W_s L_s$ and $V_b = (H_c - \delta) W_s L_s$, where the height of the interface layer $\delta$ is described by: $\delta = L_s/F = L_s D/(k_m H_c)$ outside complete delivery and $\delta = H_c$ inside complete delivery [11].

Mass conservation yields a system of ordinary differential equations (ODEs):

$$V_b \frac{dc_b}{dt} = J_{in} - J_D - J_{c_b,out} \quad (11)$$

$$V_s \frac{dc_s}{dt} = J_D - J_R - J_{c_s,out} \quad (12)$$

$$\frac{db}{dt} = k_{on} c_s (b_m - b) - k_{off} b \quad (13)$$

Equation 11 and 12 describe bulk and interface concentration evolution. Equation 13 describes Langmuir kinetics. Full derivation and implementation are detailed in **Supplementary D**. The model was implemented in Python 3.12 using *'solve_ivp'* for ODE integration.

Simulations run to user specified time, or injection time $t_{inj} = V_{in}/Q$. Post-injection time, the model operates in continuous flow ($c_{in} = 0$) or stopped flow mode. In stopped flow mode, Equation 12 reduces to:

$$V_s \frac{dc_s}{dt} = -J_R \quad (14)$$

allowing only binding/unbinding reactions.

The model outputs bulk concentration $c_b$, interface concentration $c_s$, binding rate $db/dt$, and bound density $b$ as function of time. It tracks injected and lost molecules (those exiting the system), providing insights into biosensor performance including molecule loss percentage and rate-limiting processes.

## Model validation and limitations

The two-compartment model was validated against COMSOL simulations across specific regimes (transport-limited, reaction-limited, and outside complete delivery; **Supplementary C**). Simulated binding kinetics, equilibration times, and required volumes showed strong agreement with COMSOL (RMSE < 5%, mean-relative error 7.3%), while decreasing computational time by over 100-fold.

However, four key limitations arise from the foundational well-mixed compartment assumptions:

- **Geometric bounds:** The model assumes the sensor spans the full channel width $(W_s = W_c)$ For narrower sensors $(W_s < W_c)$, target molecules laterally bypassing the sensor increases the required injection volume. Assuming negligible lateral diffusion, this volume penalty scales as $W_s/W_c$.
- **No inter-compartment transport:** Molecules cannot transport between the interfacial $(c_s)$ and bulk $(c_b)$ compartments. This approximation holds for continuous-flow systems but may omit critical diffusive transport in stopped-flow biosensors.
- **Quasi-steady state assumption:** The model requires a transport steady state to form before significant binding occurs. Consequently, the channel residency time $\tau$ must be short compared to the binding timescale $t_R$. For violations of this condition $\tau > t_R$, the well-mixed compartment assumption breaks down. In this unsteady regime, the model (and analytical solutions) can over- or underestimate the binding.
- **Formation of depletion layer:** The Sherwood approximations require a fully developed depletion layer, valid only when surface binding sufficiently depletes the target molecules: the biosensor capacity $N_{sensor} = b_{eq} W_s L_s$ must remove significant molecules from the depletion layer $N_{layer} = \delta W_c L_s c_{in}$ [11]. For $\varepsilon = \frac{N_{sensor}}{N_{layer}} \gg 1$, the Sherwood model is valid; the interface experiences reduced concentration which leads to a decreased binding rate. A limitation of our model is that this depletion constraint is considered as a binary switch: for $\varepsilon > 1$, the Sherwood approximation is valid and for $\varepsilon \leq 1$, complete delivery is considered (even for $Q > Q_c$). This switching case can occur gradually during a measurement.

## Case study 1 – Flow rate optimization for EpCAM binding

We demonstrate model application through flow rate and volume optimization for epithelial cell adhesion molecule (EpCAM) capture, a critical biomarker for multiple cancer types.

A quartz crystal microbalance (QCM) measured binding kinetics between EpCAM protein and aptamer SYL3C-10T [27]. Aptamers were co-immobilized on the gold surface via thiol-gold interactions, and proteins were injected at various flow rates. Experimentally determined parameters: $k_{on} = 4.5 \cdot 10^4\ M^{-1}s^{-1}$, $k_{off} = 7.03 \cdot 10^{-4}\ s^{-1}$ (**Supplementary A**), $b_m = 5.46 \cdot 10^{-8}\ mol/m^2$ (assuming each aptamer binds to the surface) and $D_{EpCAM} = 9.378 \cdot 10^{-10}\ m^2/s$ (Einstein approximation). The geometry was modelled with sensor area $(0.916\ cm^2)$ and chamber volume $(50\ \mu L)$, at a channel height of $0.55\ \mathrm{mm}$. Simulations with the two-compartment model and experiments were performed at $100\ nM$ to identify optimal flow rate and required volume.

Figure 4a shows equilibrium percentage across flow rates and volumes. Experimental validation (Figures 4c) demonstrates equilibrium achievement at $20\ \mu L/min$ with $200\ \mu L$. Reducing volume (Figure 4e) prevents equilibrium. Higher flow rates ($100\ \mu L/min$, Figure 4d) require larger volumes due to reaction-limited operation (Figure C.1).

Figure 4b characterized the system for equilibration time (circles) and required volume (squares) versus flow rate. COMSOL simulations (green crosses, blue plus signs; **Supplementary C**) show excellent agreement with the model. Increasing flow rate accelerates equilibration but increases volume requirements. Although the predicted critical flow rate for complete delivery is $Q \approx 20\ \mu L/min$, the 100 nM concentration yields $\epsilon < 1$, maintaining complete delivery up to Q ~ $10^3\ \mu L/min$.

This case study validates the model against experimental data and numerical simulations. Equilibration time decreases with flow rate, while required volume increases only upon entering the reaction-limited regime, demonstrating the importance of regime-aware optimization.

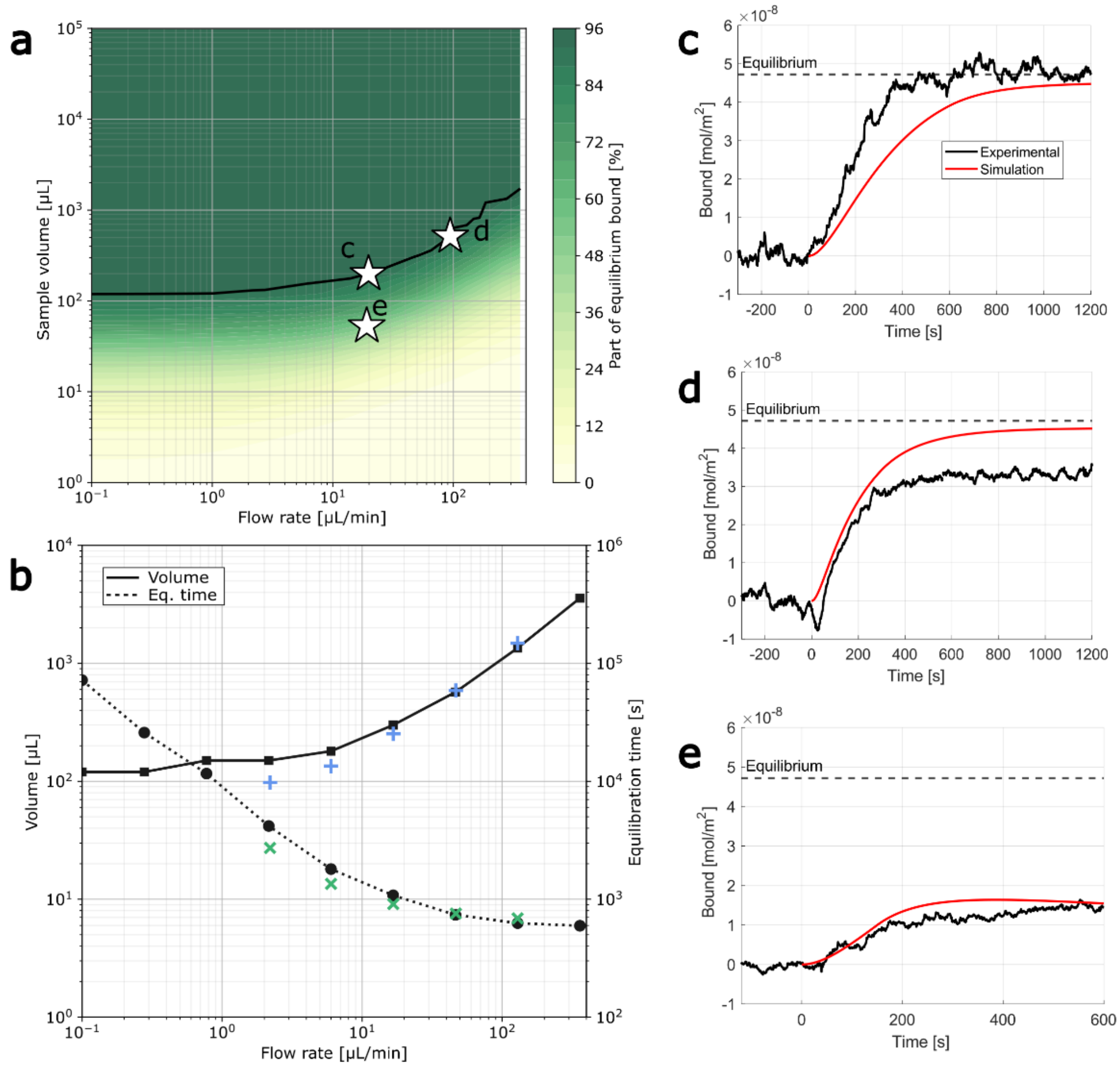

***Figure 4*** *System characteristics for EpCAM binding in a QCM sensor. (**a**) Percentage of equilibrium reached at three times the injection time. The black contour line shows the volume limit required to reach 95% equilibrium. (**b**) Minimum volume required and equilibration time for a range of flow rates. Green crosses and blue plus signs indicate the obtained equilibration time and volume from COMSOL simulations. (**c-e**) Experimental and simulation kinetics for flow rate and sample volumes cases, as indicated as white stars on 6a.*

## Equilibration time and Damköhler number

The equilibration time of a biosensor system depends on both the binding and transport of molecules. Crucially, the limiting factor of these phenomena determines the equilibration time of a system. The Damköhler number defines the ratio between the transport flux to the interface $J_D = k_m c_{in} W_s H_c$ and maximum reaction flux (at $b = 0$ and $c_s = c_{in}$) $J_R = k_{on} c_{in} b_m W_s L_s$:

$$Da = \frac{J_R}{J_D} = \frac{k_{on} b_m L_s}{k_m H_c} \quad (15)$$

Where the mass transport coefficient $k_m$ is determined by Sherwood interpolation as previously described, or can be approximated through $k_m = Q/(W_c H_c)$ inside complete delivery and $k_m \approx (D/H_c)(1.79\lambda)^{\frac{2}{3}} \cdot (Q/(DW_c))^{1/3}$ outside complete delivery regime. Thus, the flow rate $Q$ is the main operational parameter to change $Da$.

For $Da \ll 1$, the transport is much faster than the binding, and thus the system is **reaction limited**. The interface concentration approaches the input concentration ($c_s \approx c_{in}$) and the equilibration time is governed solely by binding kinetics. Conversely for $Da \gg 1$ binding is much faster than transport, and thus the system is **transport limited**. The interface is depleted ($c_s \ll c_{in}$) and equilibration depends on the transport rate. At $Da \approx 1$, the system is balanced between both processes.

In a quasi-steady state where $\frac{dc_s}{dt} \approx 0$ and at an initial unbound state ($b = 0$), the concentration of the interface is described by:

$$c_s = \frac{c_{in}}{1 + Da} \quad (16)$$

This leads to the generalized expression for equilibration time across all regimes (**Supplementary E**):

$$t_{eq} \approx 3(1 + Da)t_R \quad (17)$$

With $t_R = \left(k_{on} c_{in} + k_{off}\right)^{-1}$ and the prefactor $[-\ln(0.05)] \approx 3$ is used to convert the characteristic timescale $t_R$ to the 95% equilibrium time. This expression follows from the first order interpolation of both regimes, where for $Da \ll 1$, the equilibration time is minimized to the reaction rate $t_{eq} \approx 3t_R$. For $Da \gg 1$, the equilibration time scales linearly with the Damköhler $t_{eq} \approx 3Da \cdot t_R$. Figure 5a confirms the correspondence of our model simulations to the analytical solution of Equation 17, both inside and outside complete delivery.

## Required volume

The minimal required volume follows from the number of molecules required to saturate the surface (eq. 6): $V_{min} = (W_s L_s k_{on} b_m)/(k_{on} c_{in} + k_{off})$. The actual required volume will depend on the equilibration time and applied flowrate for a biosensor: $V_{req} = t_{eq} Q \approx [3(1 + Da)t_R]\, Q$.

**Complete delivery (Q ≤ $Q_c$).** In this regime, all molecules enter the interface compartment and are able to bind depending on binding kinetics. The relation $= k_m H_c W_c$ leads to the expression of $V_{req}$ (**Supplementary E**):

$$V_{req} \approx 3V_{min}\left(\frac{1}{Da} + 1\right) \quad \text{for } Q \leq Q_c \quad (18)$$

For $Da \gg 1$ (transport limited), the required volume approaches the minimal volume – thus indicating an efficient biosensor where all molecules bind before exiting. For $Da \ll 1$, the required volume scales as $1/Da$, as molecules pass the sensor before they can bind.

**Outisde complete delivery (Q > $Q_c$).** When the flow rate is increased past the critical flow rate $Q_c$, molecules are swept past the sensor before diffusing to the interface. This critical flow rate is given by:

$$Q_c = Pe_{H,c} W_c D \approx 1.79 \lambda W_c D \quad (19)$$

Subsequently, the critical Damköhler number $Da_c$ can be found:

$$Da_c \approx \frac{k_{on} b_m H_c}{1.79\, D} \quad (20)$$

The critical Damköhler number $Da_c$ is system dependent and determines operation outside complete delivery. Due to a loss of particles, an additional penalty for the required volume is incurred compared to biosensors in complete delivery:

$$\frac{V_{req}}{3V_{min}} \approx \left(\frac{1}{Da} + 1\right)\left(\frac{Da_c}{Da}\right)^2 \quad \text{for } Q > Q_c \quad (21)$$

Figure 5b shows the required volume as scaled to the minimum volume. Simulated points inside complete delivery (circles) follow the analytical curve of Equation 18, whereas the points outside complete delivery (squares) follow the penalized expression of Equation 21, as shown for three example cases ($Da_c = 1, 10, 100$).

### Flow rate effects

Notably, once the biosensor is outside the complete delivery regime, increasing the flow rate further ($Q > Q_c$) leads to a slower decrease in Damköhler number, as seen for the specific case $Da_c = 10$ in Figure 5c. The subsequent effect for equilibration time and required volume are shown in Figures 5d and 5e, where the hypothetical cases of complete delivery are shown in dotted lines. Notably, the analytical and simulated results show that both equilibration time and required volume deviate from the analytical result in complete delivery. However, both regimes converge, due to a decrease in Damköhler number for complete delivery ($Da \sim Q^{-1}$) compared to outside complete delivery ($Da \sim Q^{-1/3}$).

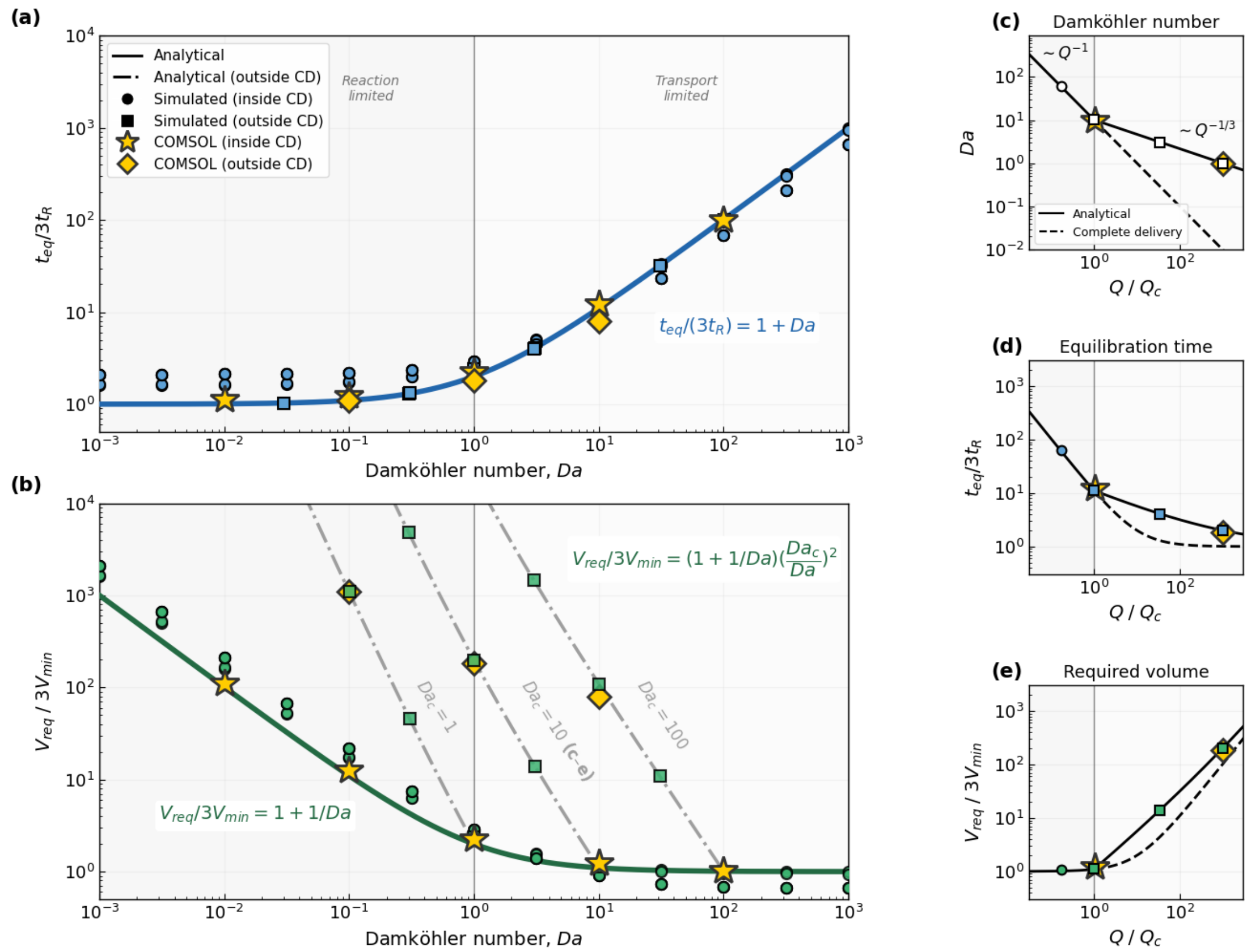

***Figure 5*** *Equilibration time and required volume for biosensors. (**a**) Equilibration time for biosensors scaled to the reaction time* $t_R$*. The equilibration time scales with the Damköhler number* $t_{eq}/3t_R = 1 + Da$*, as shown in simulations for both inside and outside complete delivery (CD). (**b**) Required volume for biosensors scaled to the minimum volume* $V_{min}$*. Inside complete delivery, the required volume is given by* $V_{req}/3V_{min} = 1 + 1/Da$*. Outside complete delivery, the required volume is given by* $V_req/3V_{min} = (1 + 1/Da)\ (Da_c/Da)$^2 *, where* $Da_c$ *is a system dependent parameter. Analytical curves for* $Da_c = 1, 10, 100$ *are shown, with corresponding matching simulation results. (**c-e**) Flow rate effects outside complete delivery for* $Da_c = 10$*. Simulations and analytical solutions collapse on the same curve and show the deviation from the (hypothetical) complete delivery regime.*

## Practical optimization guidelines

Building up on the derived analytical expressions, we obtain a practical optimization flowchart for biosensors as seen in Figure 6. This flowchart can be used to obtain simple insights into biosensor performance for both equilibration time and required volume, without the need for simulations. In general, biosensor balances two key metrics: measurement time and required volume. While increasing flow rate can accelerate equilibration, it may also increase volume requirements through particle loss. The optimization of biosensors is dependent on the specific regimes of Damköhler and complete delivery.

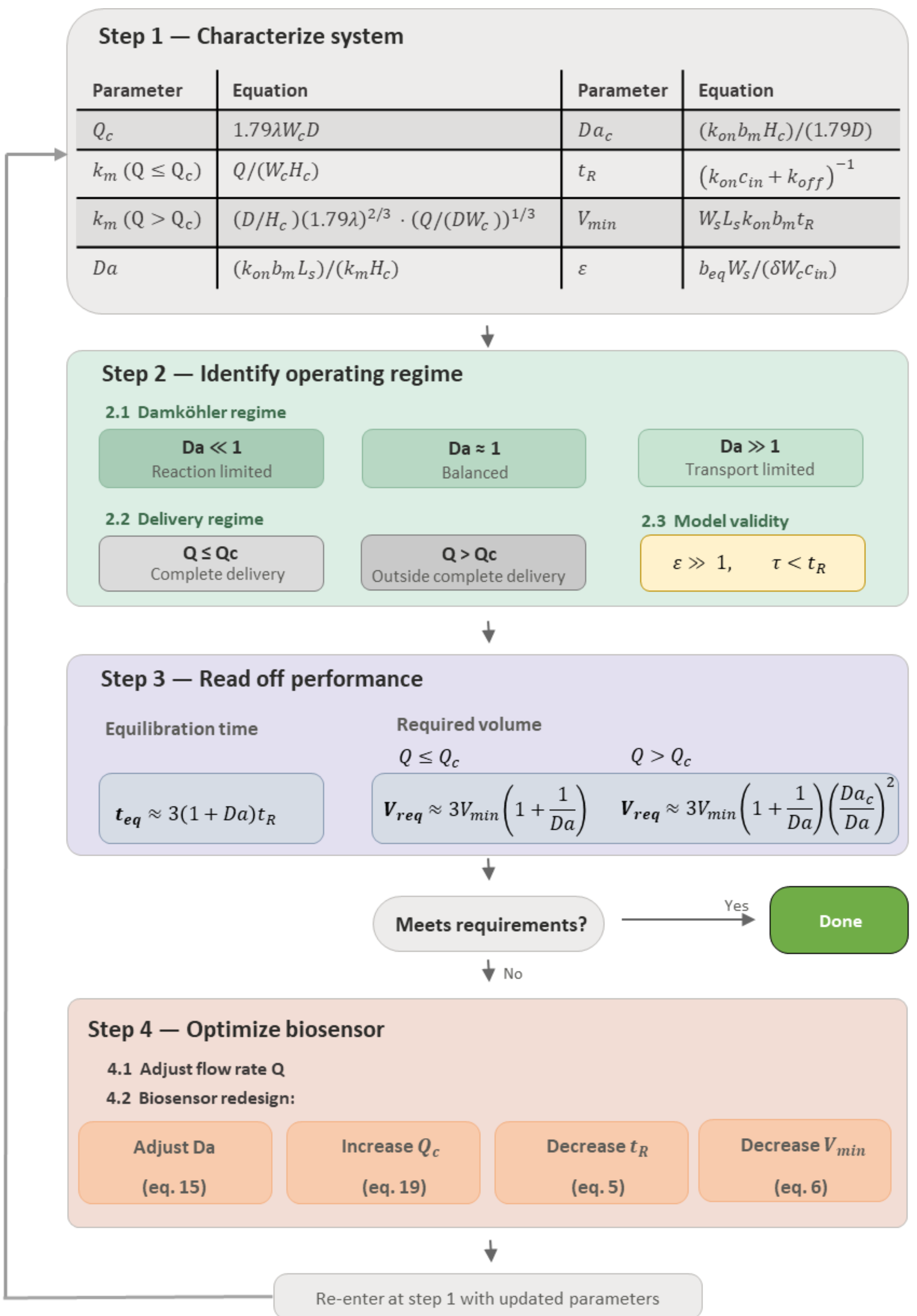

***Figure 6*** *Flowchart to optimize a biosensor with requirements ($V_{req}, t_{eq}$) or parameter constraints. The system is first characterized, and after the operating regime is identified. Accordingly, the performance can be calculated, corresponding to Figure 5. If the performance does not match the requirements, the biosensor can be optimized: mostly simply by adjusting the flow rate, or by redesigning the biosensor.*

## Optimization guidelines

- **Step 1: Characterize the system**. Obtain the system parameters from table 1c for a biosensor and calculate the described parameters. The Equation used for $k_m$ depends on the delivery regime of the biosensor, where $Q > Q_c$ indicates a system outside complete delivery.

- **Step 2: Identify the operating regime.** The Damköhler number places the system on a continuous spectrum from reaction limited (Da $\ll 1$) through balanced (Da $\approx 1$) to transport limited (Da $\gg 1$). Simultaneously, the flow rate relative to $Q_c$ determines whether the biosensor operates in complete delivery (Q ≤ $Q_c$) or outside (Q > $Q_c$). These regimes are visualized along the Da axis of Figures 5a and 5b. Additionally, $\varepsilon$ and $t_R$ determine model validity. For $\varepsilon < 1$, a complete delivery can be used (even for $Q > Q_c$). For $\tau > t_R$, the model shows deviations from numerical solutions due to well-mixed assumptions.
- **Step 3: Read off performance.** The equilibration time is given by $t_{eq} \approx 3(1 + Da)t_R$. The required volume follows from $V_{req} \approx V_{min}(1 + 1/Da)$ inside complete delivery, with an additional penalty factor $(Da_c/Da)^2$ outside complete delivery. These performance metrics are compared against the available sample volume and measurement time.
- **Step 4: Optimize.** If performance is insufficient, the biosensor can be optimized:
  - The flow rate can be adjusted, which will move the predicted $t_{eq}$ and $V_{req}$ according to the analytical solutions as shown in Figures 5a and 5b.
  - The biosensor can be redesigned to adjust the key intermediate variables. Adjustment of $Da$ or $Q_c$ leads to a move of predicted $t_{eq}$ and $V_{req}$ according to the analytical solutions in Figures 5a and 5b. Adjustment of $t_R$ and $V_{min}$ leads to a vertical shift for these curves.

In summary, understanding biosensor operating regimes is essential for optimization: first, to identify current system limits; second, to guide parameter modifications for regime improvement. The interplay between equilibration time and volume, constrained by available sample volume, determines optimization boundaries.

## Case study 2 – Optimization of a PSA biosensor

We demonstrate the use of the biosensor flowchart from Figure 6 on a free prostate-specific antigen (PSA) detection platform [28] with linear detection from 21.4 – 1000 ng/mL (0.82–38.3 nM, MW = 26,079 Da).

*Step 1: Characterize.* From the geometry ($W_c = 200\ \mu m,\ L_c = 3\ mm, H_c = 20\ \mu m$), sensor parameters ($W_s = 200\ \mu m, L_s = 3\ mm, b_m = 1.5 \cdot 10^{-7}\ mol\ m^{-2}\ k_{on} = 2.54 \cdot 10^5\ M^{-1}s^{-1}$, $k_{off} = 4.45 \cdot 10^{-3}\ s^{-1}$) and sample parameters ($D = 8.5 \cdot 10^{-11}\ m^2s^{-1}, c_{in} = 0.38\ nM, Q = 0.5\ \mu L/min$), the biosensor is characterized. This obtains $Q_c = 0.27\ \mu L/min$, $k_m = 0.0014\ m/s$, $Da = 4.10, Da_c = 5.01, t_R = 220\ s, V_{min} = 15.1\ \mu L$ and $\varepsilon = 625$.

*Step 2: Identify operating regime.* A Damköhler number of 4.10 signifies a balanced but slightly transport limited system. As $Q > Q_c$, the biosensor operates outside the complete delivery regime.

*Step 3: Read off performance.* The predicted equilibration time $t_{eq} \approx 3363\ s\ (\sim 56\ min)$. The required volume is calculated through the outside complete delivery Equation, obtaining $V_{req} \approx 28.0\ \mu L$, which is beyond the available $5\ \mu L$. Therefore, the system only reaches a part of the equilibrium, limiting sensitivity.

*Step 4: Optimization.*

1. *Adjust flow rate.* Most simply, the flow rate can be reduced to match the critical flow rate $Q_c = 0.27\ \mu L/min$, which places the system back in the complete delivery regime, but increases the $Da = 5.01$. Subsequently, the equilibration time is increased to $t_{eq} \approx 3964\ s\ (\sim 66\ min)$, but the required volume is decreased to $V_{req} \approx 18.1\ \mu L$. The improvement in volume requirement follows from a higher Damköhler number, as well as remaining inside complete delivery. Further decreasing the flow rate increases the Damköhler number further, leading to $V_{req} \rightarrow V_{min}$.
2. *Biosensor redesign.*
    - *Adjustment of $b_m$.* The minimal volume $V_{min}$ can be adjusted by decreasing the number of probes to $b_m = 1.5 \cdot 10^{-8}\ mol\ m^{-2}$, which for a flow rate of $0.27\ \mu L/min$ leads to a $V_{min} = 1.5\ \mu L$, a Damköhler number of 0.41, $V_{req} \approx 4.5\ \mu L$ and $t_{eq} \approx 990\ s\ (\sim 16.5\ \mathrm{min})$. However, this leads to a lower signal due decreased probe density.
    - *Adjustment of $H_c$.* $Da$ and $Q_c$ can be adjusted by reducing the channel height to $H_c = 10\ \mu m$. With the changed geometry, the critical flowrate is increased to $Q_c = 0.55\ \mu L/min$, thus the system remains in complete delivery for Q = 0.50 $\mu L/min$. Additionally, the Damköhler decreases to 2.74 due to more effective transport to the surface, which leads to $t_{eq} \approx 2470\ (\sim 41\ min)$ and $V_{req} \approx 20.6\ \mu L$. The equilibration time is improved due to a lower $Da$, and the required volume is improved by remaining in complete delivery.

Figure 7 shows the flow rate effects on the described biosensor, both for the original design (in black) and for the described redesigns (green and orange). Notably, the original design with a flow rate of $Q = 0.5\ \mu L/min$ leads to a higher volume requirement than the available $5\ \mu L$. Therefore, improvements in biosensor design or supplied volume would lead to higher bound molecule density, which would increase biosensor sensitivity.

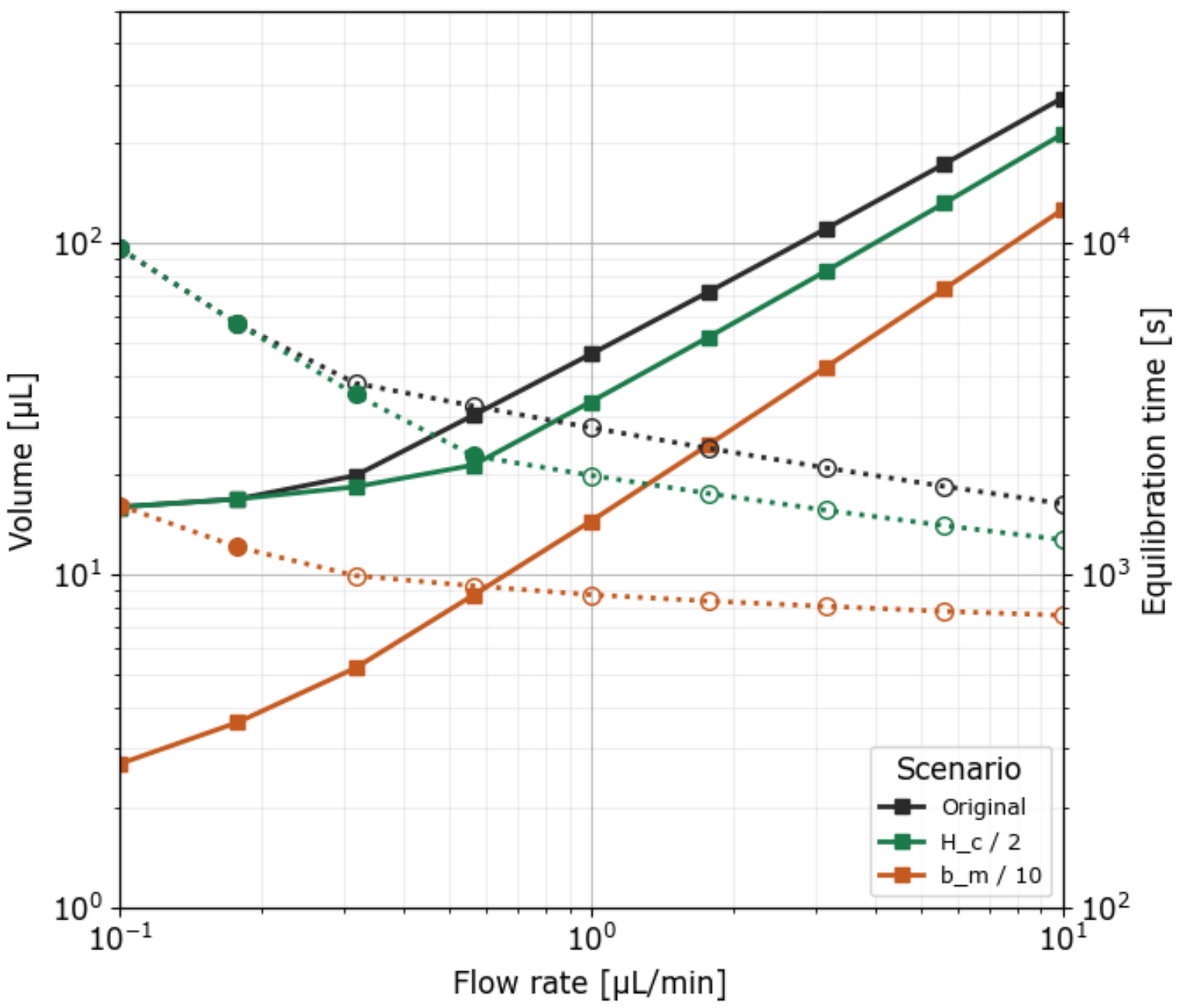

***Figure 7*** *System characteristics for the PSA biosensor. Minimum volume required (squares) and equilibration time (circles) for a range of flow rates. Filled circles indicate measurement points in the complete delivery regime. Points in black showcase flow rate optimizations done for the parameters of the original biosensors. Points in green and orange showcase the biosensor redesign implications. A shorter channel height leads to a shift in the critical flow rate, thus allowing to the biosensor to remain in complete delivery for higher flow rates. A decreased probe density leads to improvements in both* $t_{eq}$ *and* $V_{req}$*, at the cost of decreased signal.*

## Conclusions

This work addresses a fundamental yet overlooked constraint in affinity-based biosensing that directly impacts clinical utility: the finite volume of real-world samples. While transport phenomena and binding kinetics are well studied, biosensor performance ultimately depends on the absolute number of target molecules available for capture, determined jointly by concentration and sample volume. By incorporating volume limitations into biosensor modeling, we provide a framework that bridges laboratory performance to clinically relevant conditions.

We developed a computationally efficient two-compartment model combining simplified mass transport with Langmuir binding kinetics under finite-volume conditions. The model shows good correspondence with COMSOL simulations for binding curves (root-mean-square error < 5%) and predicted equilibration time and required volume (mean relative error of 7.3%), while achieving a >100-fold reduction in computational time, enabling rapid design iteration and parameter exploration.

Additionally, we derived analytical expressions for equilibration time and minimum required sample volume as a function of the Damköhler number, valid across transport-limited and reaction-limited systems. We identify the system dependent critical flow rate $Q_c$ and corresponding critical Damköhler number $Da_c$ as a determination of the complete delivery regime in which all input target molecules can interact with the biosensor. Crucially, biosensors outside this regime require additional sample volume. Analytical solutions showed good agreement with COMSOL for predicted equilibration time and volume requirement (mean relative error ~ 11%).

These results reframe biosensor optimization around clinically relevant constraints of volume availability and measurement time, demonstrated through a practical optimization guide and application to a PSA biosensor. The open-source implementation ensures broad accessibility, equipping researchers to rapidly explore design spaces and predict performance under realistic conditions, accelerating the translation of biosensors from laboratory settings to real-world clinical applications.

# Acknowledgements

Figures created with https://BioRender.com. We would like to acknowledge funding from the Swiss National Science Foundation (SNSF) through the Health and Wellbeing Call, project HYPERCELL (31HW-0_220646).

# Code availability

The open-source biosensor model is available on GitHub: https://github.com/dgbeijersbergen/biosensor_model.

# Credits:

DB: Conceptualization, Data curation, Formal analysis, Investigation, Methodology, Resources, Software, Validation, Visualization, Writing – original draft, Writing – review and editing. JC: Conceptualization, Formal analysis, Funding acquisition, Methodology, Resources, Supervision, Visualization, Writing – review and editing.

# Sample volume as a key design parameter in affinity-based biosensors - Supplementary Material

Daan Beijersbergen[1,2], Jérôme Charmet[1,2]

Correspondence to: jerome.charmet@he-arc.ch

3. School of Engineering–HE-Arc Ingénierie, HES-SO University of Applied Sciences Western Switzerland, Neuchâtel, Switzerland
4. School of Precision and Biomedical Engineering, University of Bern, Bern, Switzerland

# Supplementary A: experimental section

### Methods

The binding affinity was determined by quartz crystal microbalance (QCM – OpenQCM-Q1, Novaetech S.r.l) binding between EpCAM antigens and anti-EpCAM aptamers. Initially, the QCM was calibrated for initiation and temperature shifts. Afterwards, a PBS buffer was injected into the QCM chamber, and at least 15 minutes waited for an equilibrium. The surface was subsequently functionalized with aptamers, before the target molecule EpCAM was injected.

*Materials*

Aptamers (single strand oligonucleotides) were ordered from Microsynth (Balgach, Switzerland) with the sequence 5'-CAC TAC AGA GGT TGC GTC TGT CCC ACG TTG TCA TGG GGG GTT GGC CTG TTT TTT TTT T-3' with a thiol bond on the 3' end. Liquid sensing QCM sensors were ordered from OpenQCM (Pompeii, Italy). 6-Mercaptohexanol (MCH) and tris(2-carboxyethyl) phosphine (TCEP) were ordered from Sigma-Aldrich (Buchs, Switzerland). Epithelical Cell Adhesion Molecule (EpCAM) protein (MW = 31-40 kDa) was ordered from AcroBiosystems (Zug, Switzerland).

*Surface functionalization*

The gold surface of the QCM was functionalized with aptamers following standardized co-immobilization with 6-mecaptohexanol (MCH) [1]. In brief, thiolated aptamers were first re-suspended in a $10\,\mu M$ solution of PBS + 5 mM $Mg^{2+}$ (hereafter: PBS+). The solution was reduced for 30 minutes in a TCEP solution, where the ratio of aptamer to TCEP was 1:100. The reduced solution was then diluted in PBS+ to a concentration of 200 nM. The solution was heated to $90^o\ C$ for 5 minutes and was allowed to cool down to room temperature over 15 minutes, to allow for optimal conformation of the aptamer [2]. Simultaneously, fresh MCH was diluted in PBS+ to a concentration of 40 $\mu M$. MCH and aptamer solution were combined in a 1:1 (v/v) solution and immediately injected to functionalize the surface, during a period of 15 minutes. The surface was subsequently washed with PBS+ and backfilled with $40\ \mu M$ MCH solution. The functionalized surface was utilized immediately after.

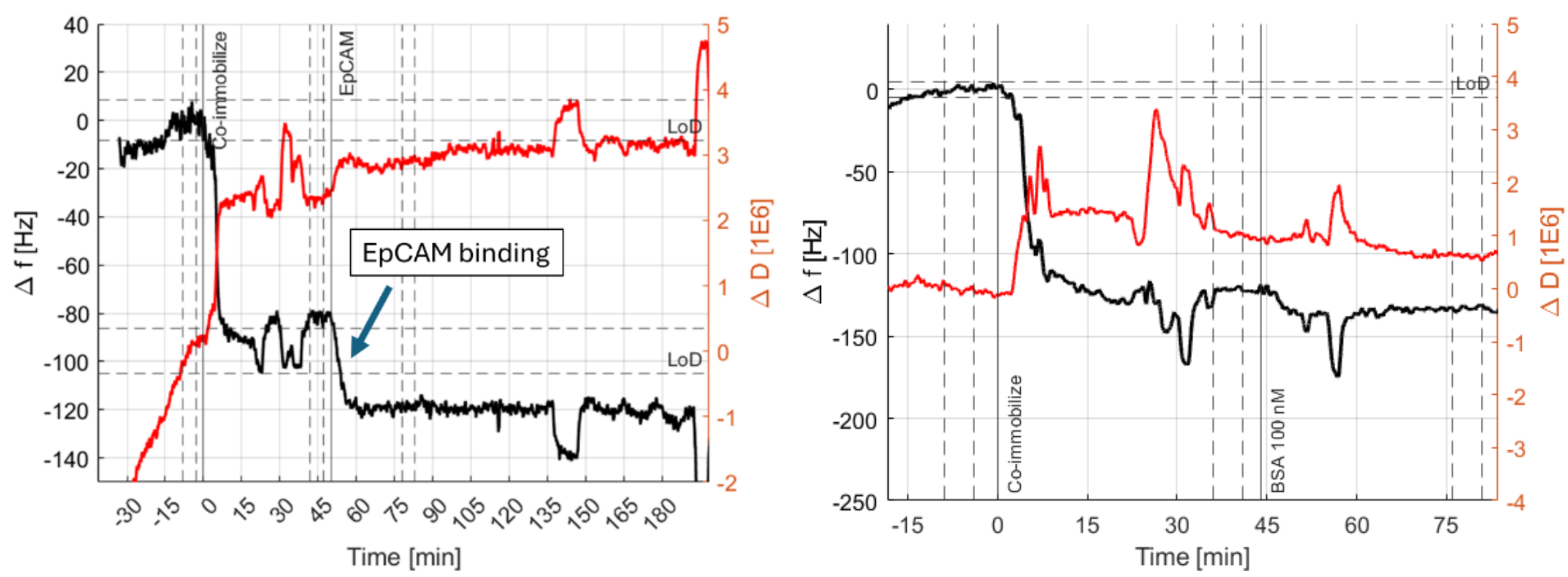

*Figure A.1 – **(a)** Example QCM measurement data. PBS+ was injected and subsequently the frequency was stabilized. At t = 0, the $50\ \mu L$ co-immobilization solution with 100 nM aptamers and 40 $\mu M$ MCH was injected, which led to a clear shift in frequency and an increase in dissipation. Subsequent washing in PBS+ (t = 20 min), backfilling (25 min) and washing in PBS+ (t = 30 min) did not lead to a bigger frequency shift, thus indicating a fully functionalized surface. After a short equilibrium period, the EpCAM solution of 100 nM was injected at 20 $\mu L/min$. Stabilization of the signal after the binding event indicates a strong bind and potential re-binding events. **(b)** Control experiment with BSA. The same protocol was used as in Figure A1a with 100 nM BSA instead of 100 nM EpCAM. The frequency shift was -11.3 Hz BSA compared to -37.3 Hz for EpCAM.*

An example of QCM binding kinetics can be observed in Figure A.1a. Here, EpCAM protein was injected at t = 50 min, which induces a frequency shift of $37.3\ Hz$. As a control, the same protocol was repeated with BSA protein in the same concentration of 100 nM, as seen in Figure A. 1b. Injection of BSA at t = 45 leads to a smaller shift of $11.3\ Hz$, indicating some non-specific binding.

In the Sauerbrey equation, the bound mass to the surface is related to the frequency shift: [3]

$$\Delta f = -C_f \Delta m \text{ (A.1)}$$

With $\Delta f$ frequency shift $(Hz)$, $C_f$ the mass sensitivity $(g\ Hz^{-1} cm^{-2})$ and $\Delta m$ the change in mass to the surface $(g)$. Given the $C_f$ for the QCM of $4.42 \cdot 10^{-9}\ g\ Hz^{-1} cm^{-2}$, the surface area of the QCM of $0.916\ cm^2$ and the average molecular weight of EpCAM of 35.5 g/mol after glycosylation, the frequency shift was converted in a bound density to the surface. It should be noted that the conversion as given in Equation A.1 was originally established for thin and rigid layers, and that the utilization of this equation for the binding of soft biomolecules presents limitations. However, it has been shown that Equation A.1 is valid for small proteins below 150 kDa, therefore the frequency shift caused by EpCAM binding (MW = 31 – 40 kDa) can be correlated to the bound mass [4].

The bound equilibrium was measured at $4.72 \cdot 10^{-8}\ mol/m^2$ (triplicate measurement). Given a functionalized surface with a density of $b_m = 5.46 \cdot 10^{-8}\ mol/m^2$, this yields a $k_D = 15.7 \pm 10.6\ nM$, which is consistent with literature for this EpCAM – aptamer interaction [5].

The $k_{on}$ and $k_{off}$ values were then determined experimentally by analyzing the binding kinetics at sufficiently high flow rates ($> 100\ \mu L$) such that the system is reaction limited ($Da < 1$) (Figure C.1a).

A problem with using QCM at such high flow rates is that the increased pressure induced a (temporary) frequency shift. At flow rates $\leq 250\ \mu L/min$, the frequency shift at injection is repeatable. Injection of phosphate buffered saline (PBS) was thus repeated three times, to obtain a mean pulse (Figure A1). This pulse was then used to compensate for the original experiment, from which binding kinetics could be obtained (Figure A2).

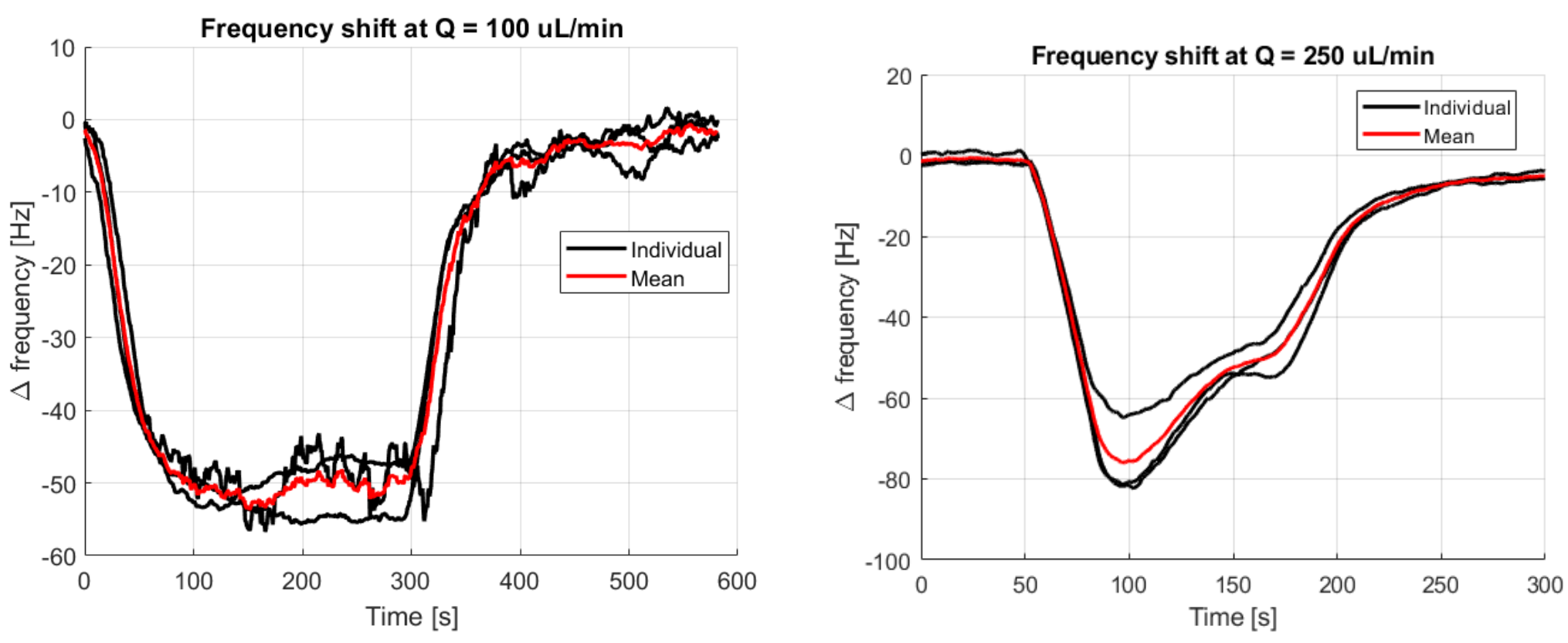

*Figure A.2 – QCM frequency shift due to injection. Black lines show individual pressure shifts after injection of PBS. The red line shows the mean of these measurements.*

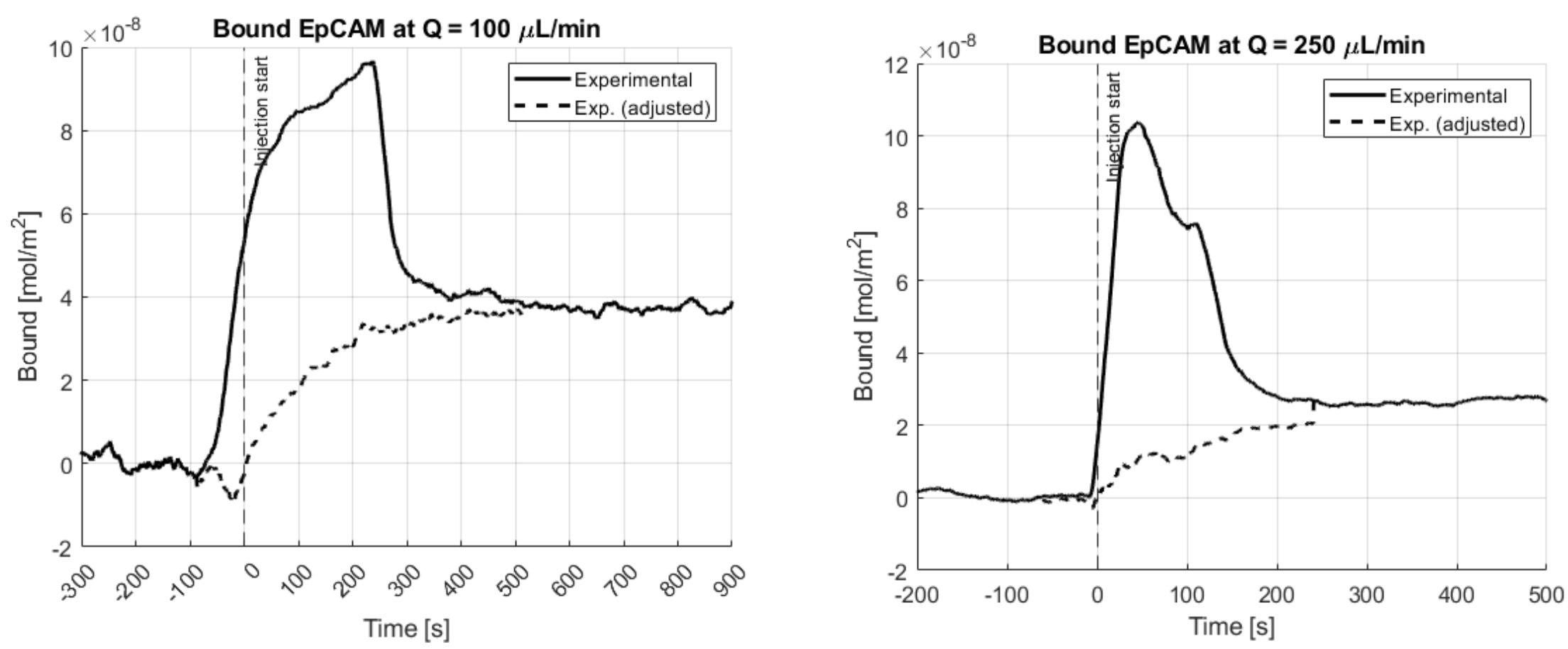

A

*Figure A.3 – EpCAM binding kinetics. The plot shows the density of bound EpCAM, obtained from QCM frequency shift and the Sauerbrey relation. The solid line shows the raw experimental data, whereas the dotted line shows the data after pressure pulse compensation.*

Langmuir model fitting the curve of Equation 4 in the main text was then simulated for a range of values for $k_{on}$ and $k_{off}$. The accuracy of the fit was determined by root mean square error, where a fit curve $X$ that has a good fit compared to the original curve $x_0$ has the smallest RMSE value:

$$RMSE = \sqrt{\frac{1}{n}\sum_{i=1}^{n}(X_i - x_0)^2} \text{ (eq. A1)}$$

However, this led to a wide range of combinations in $k_{on}$ and $k_{off}$. This was simplified by limiting the available possibilities to a range where the ratio $\frac{k_{off}}{k_{on}} = k_D$ is within 0.5 nM. This led to the $k_{on}$ and $k_{off}$ values shown in Figure A.4, along with the best fits. It should be noted that the dataset for Figures A.4a and A.4b are the same as for Figures 4c and 4d. For the determination of the binding affinity, only the Langmuir binding model was used (Equation 4 in the main text), thus re-using the dataset for the model was deemed valid. For the binding affinity, the binding curve was translated by one residence time $\tau = V/Q$ to account for filling of the QCM sensor volume, thus further limiting transport effects.

Experimental binding curves from figures 4c, 4d and 4e from the main text were compared to our python model with normalized root-mean square error calculation (equation C.1). The relative errors were 16.6%, 22.0% and 6.2% respectively. The errors between experimental results and our python model simulations are higher than the errors between our model and numerical COMSOL simulations (figure C.3), which is explained by geometry simplifications, uncertainty in bound probe density $b_m$ and general experimental variation.

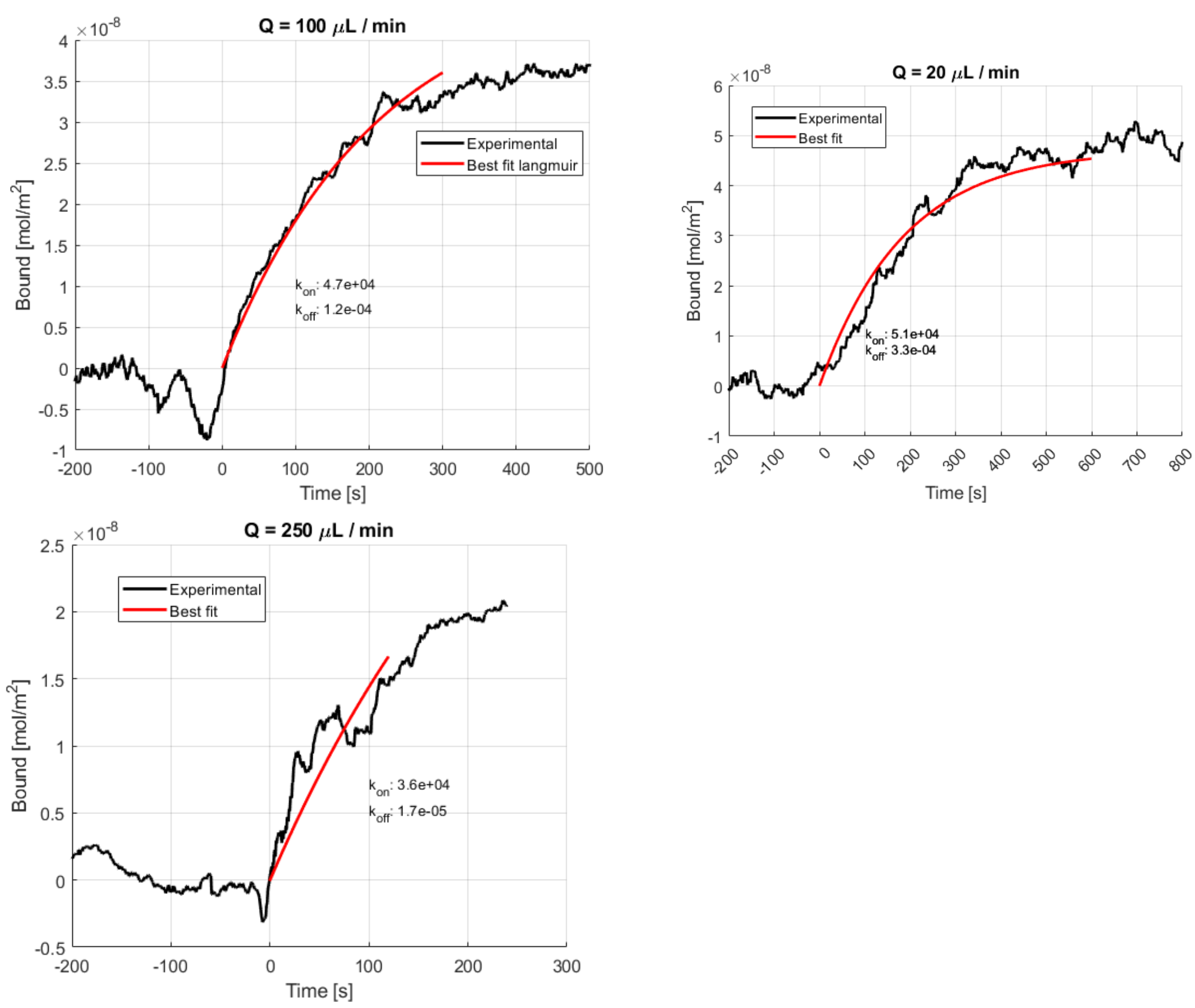

***Figure A.4*** *Langmuir fitting (red) to QCM experimental data (black).*

# Supplementary B: Sherwood number

Given the three boundary fluxes given by complete delivery, Ackerberg and Newman, an approximation of the function $F(Pe_H, Pe_s)$ is obtained through blending of the latter two limits:

$$\log_{10}(F_{eff}) = (1-\Omega)\log_{10}\left(F_{Ackerberg}\right) + \Omega\log_{10}(F_{Newman})$$

$$F_{eff} = F_A^{(1-\Omega)} \cdot F_N^{\Omega}$$

With $\Omega$ a hyperbolic tangent in range $\Omega = [0,1]$:

$$\Omega = 0.5\left(1 + \tanh\left(\alpha\,(\kappa - 0.5)\right)\right)$$

With $\alpha$ the sharpness of the function and $\kappa$ the linearized and normalized $Pe_s$ value (such that the function can be scaled to the linear domain of the hyperbolic tangent):

$$\kappa = \frac{\log_{10}(Pe_s) - \log_{10}(Pe_s, low)}{\log_{10}(Pe_s, high) - \log_{10}(Pe_s, low)}$$

Figure B.1 shows the effect of different tangent sharpness values $\alpha$. A sharpness value of $\alpha = 4$ was used as it gave the most comparable results to Squires et al. [6], without noticeable inconsistencies for the plotted values of $\lambda$. The boundary flux of full capture is enforced by limiting the flux such that $F \leq Pe_h$.

The resulting obtained Sherwood numbers in Figure 3a in the main text is akin the figure layout is akin to Figure 3b in Squires et al. (which was obtained through numerical simulations) [6].

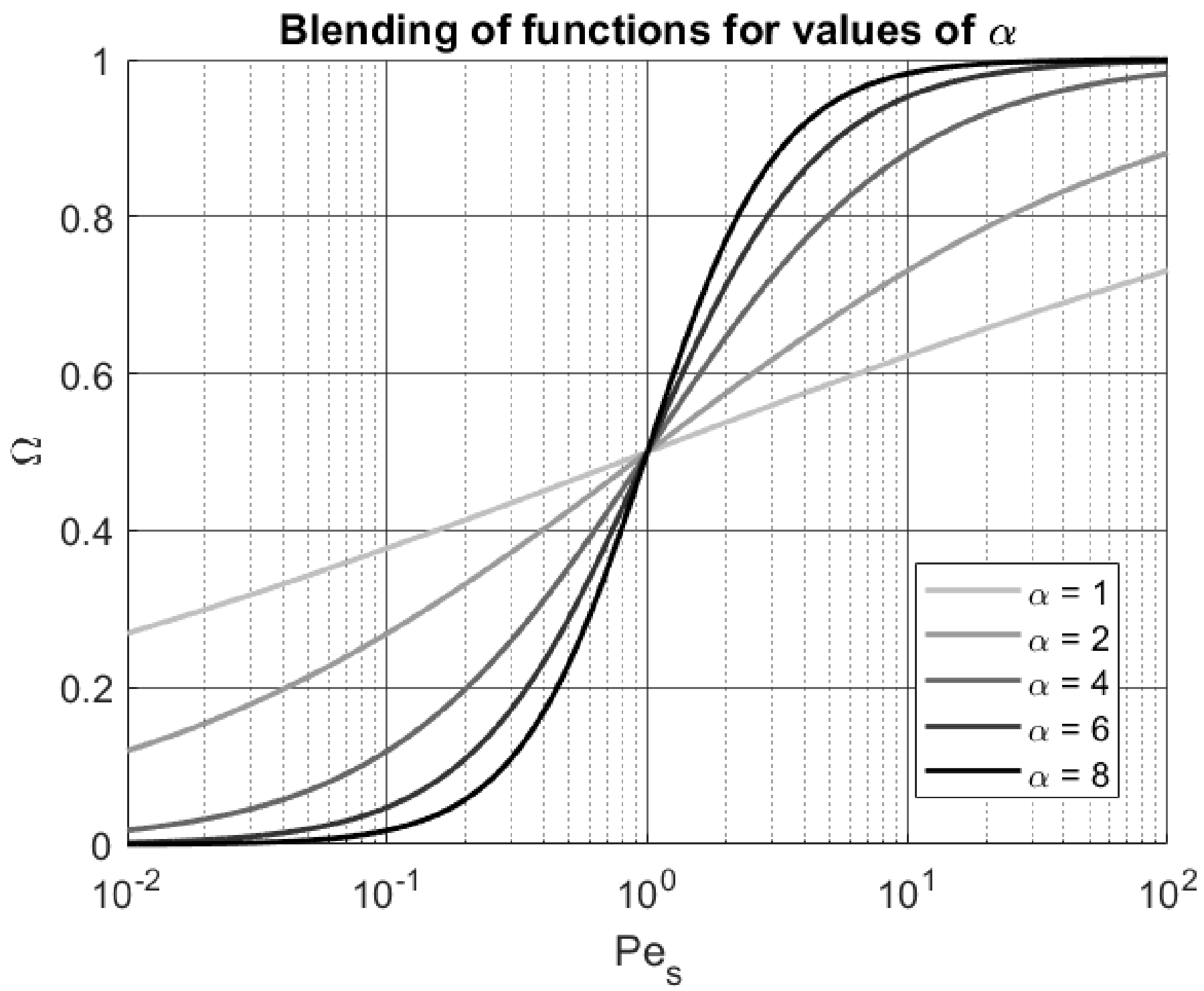

***Figure B.1*** *Various blending functions for values of* $Pe_s$.

A sensitivity analysis was performed to investigate the effect of the sharpness value $\alpha$. A biosensor outside complete delivery regime ($Q > Q_c$) was simulated in the blending region ($Pe_s \epsilon [0.01,100]$) for $\lambda\ \epsilon\ [0.1,10]$ and $Pe_H \epsilon [0.1,100]$. Figure B.2 shows the effect of sharpness values $\alpha$ on the Sherwood number, equilibration time and required volume of a biosensors. Relative small changes are observed for $\alpha \geq 3$, with variations for $|\Delta F| \leq 5\%$, $|\Delta t_{eq}| \leq 1\%$ and $|\Delta V_{req}| \leq 1\%$. This indicates that the model is largely insensitive for changes in sharpness value $\alpha \geq 3$.

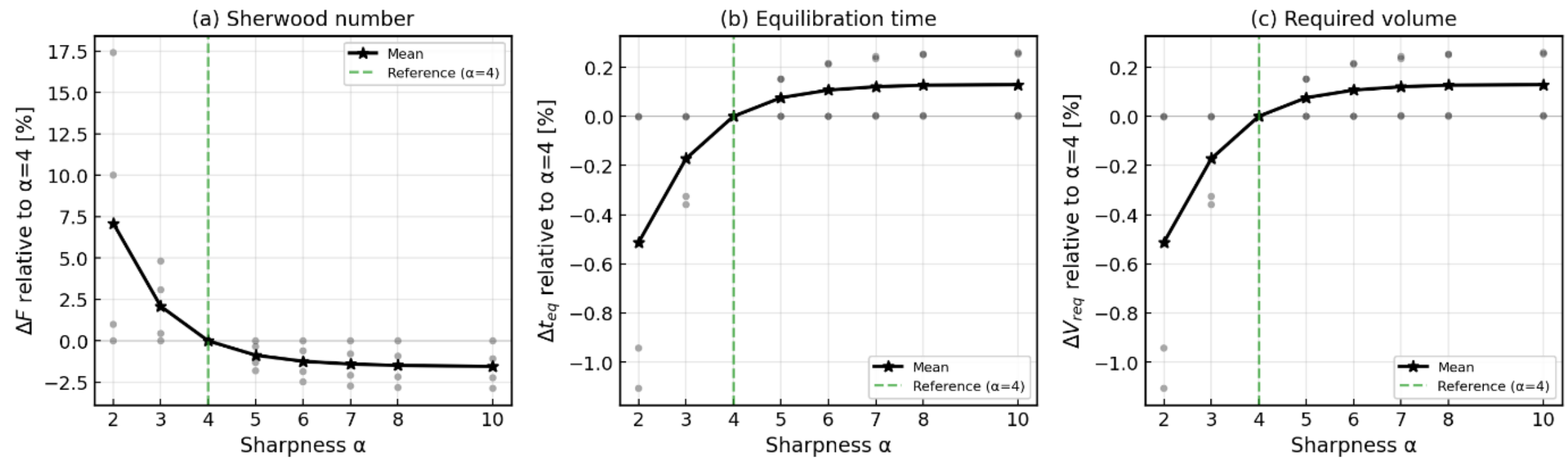

***Figure B.2*** *Sensitivity analysis for a range of sharpness values* $\alpha$.

# Supplementary C: COMSOL simulations

The biosensor model as described in the main text was validated with numerical simulations. For this purpose, COMSOL Multiphysics 6.2 was utilized. The module 'transport of diluted species' was used to numerically simulate the transport of the target molecules (including convection and diffusion). To model the binding, a general form boundary PDE was employed:

$$\frac{db}{dt} = k_{on}c(b_m - b) - k_{off}b$$

A time-dependent solution was computed to obtain kinetics for the COMSOL model. The input velocity was obtained by:

$$v_{avr} = \frac{Q}{A}$$

With $Q$: flow rate ($m^3/s$) and $A = W_c H_c$: channel cross-section ($m^2$).

Mesh triangulation was performed with the predefined 'normal' size, with a maximum element size of $1\ \mu m$ at the biosensor boundary to ensure high precision at the depletion layer.

## Box 1

COMSOL simulations for box 1 were performed with an implicit solver (BDF), free time stepping with a maximum simulation time of 5000 seconds. The parameters used for the COMSOL simulations are described in table C.1 and are identical to simulations with our two-compartment model.

| **Type** | **Parameter** | **Value** | **Unit** |
|---|---|---|---|
| Device | $W_c$ | $0.9571 \cdot 10^{-2}$ | $m$ |
| | $L_c$ | $0.9571 \cdot 10^{-2}$ | $m$ |
| | $H_c$ | $5.4579 \cdot 10^{-4}$ | $m$ |
| Sensor | $W_s$ | $0.9571 \cdot 10^{-2}$ | $m$ |
| | $L_s$ | $0.9571 \cdot 10^{-2}$ | $m$ |
| | $b_m$ | $5.46 \cdot 10^{-8}$ | $mol/m^2$ |
| | $k_{on}$ | $4.5 \cdot 10^4$ | $M^{-1}s^{-1}$ |
| | $k_{off}$ | $7.03 \cdot 10^{-4}$ | $s^{-1}$ |
| Sample | $D$ | $0.9378 \cdot 10^{-9}$ | $m^2/s$ |
| | $c_{in}$ | $100$ | $nM$ |
| | $c_0$ | $0$ | $M$ |
| | $V_{in}$ | $-$ | $m^3$ |
| | $Q$ | $variable$ | $m^3/s$ |

***Table C.1*** *Simulation parameters for COMSOL simulations of box 1*

## Validation optimization regimes

Eight points from figure 5a and 5b were picked as representative points for comparison with COMSOL. These points were chosen specifically, such that the different regimes could be compared (low, medium and high Da; within and without full collection). For these simulations, biosensor parameters as described in table C.2 were used. All simulations were performed in the regimes where our python model is valid: $\varepsilon \gg 1$ and $\tau < t_R$.

| Type | Parameter | Value | Unit |
|---|---|---|---|
| Device | $W_c$ | $1.0 \cdot 10^{-4}$ | $m$ |
| | $L_c$ | $1.0 \cdot 10^{-3}$ | $m$ |
| | $H_c$ | $1.0 \cdot 10^{-4}$ | $m$ |
| Sensor | $W_S$ | $1.0 \cdot 10^{-4}$ | $m$ |
| | $L_s$ | $1.0 \cdot 10^{-3}$ | $m$ |
| | $b_m$ | $1.0 \cdot 10^{-7}$ | $mol/m^2$ |
| | $k_{on}$ | $variable$ | $M^{-1}s^{-1}$ |
| | $k_{off}$ | $1.0 \cdot 10^{-3}$ | $s^{-1}$ |
| Sample | $D$ | $1.0 \cdot 10^{-11}$ | $m^2/s$ |
| | $c_{in}$ | $1.0$ | $nM$ |
| | $c_0$ | $0$ | $M$ |
| | $V_{in}$ | $-$ | $m^3$ |
| | $Q$ | $variable$ | $m^3/s$ |
| Characteristics | $Q_c$ | $0.0011$ | $\mu L/min$ |

***Table C.2*** *Simulation parameters for COMSOL simulations of figure 5.*

For the biosensor with parameters of table C.2, the critical flow rate is $Q_c = 0.0011\ \mu L/min$. Table C.3 shows the values of $k_{on}$ and $Q$ used to obtain the desired simulations points inside complete delivery ($Da \approx 0.01, 0.1, 1, 10, 100$) and outside complete delivery ($Da \approx 0.1, 1, 10$). The simulation time for both python and COMSOL simulations were varied depending on the parameters.

| ID | $\mathbf{Q \leq Q_c}$ | **Q** $[\mu L/\mathrm{min}]$ | $k_{on}$ $[M^{-1}s^{-1}]$ | Da | $b_{eq}$ $[mol/m^2]$ | Max time (s) |
|---|---|---|---|---|---|---|
| 1 | Yes | $0.0011$ | $18$ | $0.01$ | $1.8 \cdot 10^{-12}$ | $1.0 \cdot 10^4$ |
| 2 | Yes | $0.0011$ | $1.8 \cdot 10^3$ | $0.99$ | $1.78 \cdot 10^{-10}$ | $1.0 \cdot 10^4$ |
| 3 | Yes | $0.0011$ | $1.8 \cdot 10^4$ | $9.91$ | $1.77 \cdot 10^{-9}$ | $5.0 \cdot 10^4$ |
| 5 | No ($Da_c = 10$) | 1.098 | $1.8 \cdot 10^4$ | 1.0 | $1.77 \cdot 10^{-9}$ | $1.0 \cdot 10^4$ |
| 6 | No ($Da_c = 100$) | $1.0740$ | $1.79 \cdot 10^5$ | 10 | $1.52 \cdot 10^{-8}$ | $1.0 \cdot 10^5$ |
| 7 | Yes | $0.0011$ | $1.8 \cdot 10^2$ | 0.1 | $1.8 \cdot 10^{-11}$ | $1.0 \cdot 10^4$ |
| 8 | No ($Da_c = 1$) | $1.0740$ | $1.79 \cdot 10^3$ | 0.1 | $1.79 \cdot 10^{-10}$ | $1.0 \cdot 10^4$ |
| 11 | Yes | $0.0011$ | $1.8 \cdot 10^5$ | $99.1$ | $3.2 \cdot 10^{-9}$ | $4.0 \cdot 10^5$ |

***Table C.3*** *Simulation characteristics.*

The simulated geometry with 'normal' mesh refinement and maximum element size of $1\ \mu m$ around the sensor leads to a simulation in COMSOL with 25469 degrees of freedom. Solving in COMSOL was performed with free time stepping and limited time stepping. For limited stepping COMSOL simulations, the time step constraint is given by : max(1, min(10000, 0.01*t)). This constraint leads to a maximum timestep of 1s for the first 100 seconds, and grows with 0.01 consecutive timesteps. Limited time-stepping allows more accurate kinetical results.

Table C.4 and figure C.1 show the simulation times for our python model and the COMSOL model. The COMSOL simulations in free time stepping model were on average ~409x slower, and ~2578x slower in limited time stepping. Notably, the degrees of freedom for this simulation were still small, and larger or more complex biosensors would require longer simulation time. In contrast, our python model depends on a system of ODEs with simplified transport, thus the simulation time is much shorter even for bigger geometries.

Simulations were performed on a computer with the following specifications: Processor: 13th Gen Intel Core i7-1355U, 1700 Mhz, 10 Cores; RAM: 32 GB.

| ID | Python model | COMSOL (free time stepping) | COMSOL (limited time stepping) |
|---|---|---|---|
| 1 | 0.40 | 174 | 1428 |
| 2 | 0.34 | 59 | 635 |
| 3 | 0.38 | 128 | 930 |
| 5 | 0.28 | 89 | 942 |
| 6 | 0.39 | 280 | 1506 |
| 7 | 0.38 | 42 | 710 |
| 8 | 0.41 | 140 | 746 |
| 11 | 0.82 | 690 | 1494 |

***Table C.4*** *Simulation time in seconds.*

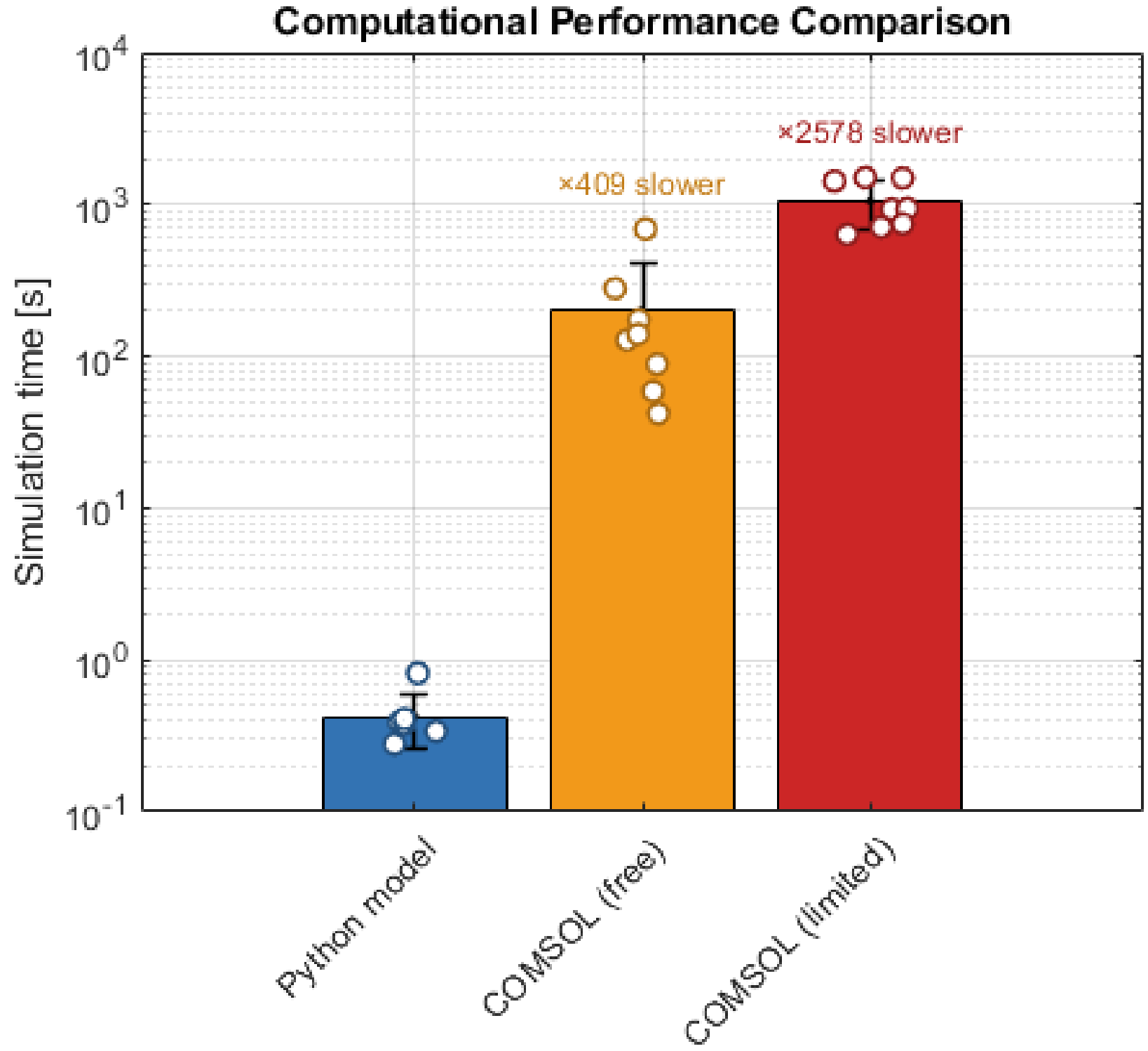

***Figure C.1*** *Simulation time comparison.*

The binding kinetics were extracted for our python model and the (limited time stepping) COMSOL simulations. Both the bound molecule density $[mol/m^2]$ and binding rate $[mol/(m^2s)]$ are shown in figure C.2, superimposed on the (theoretical maximum) Langmuir binding.

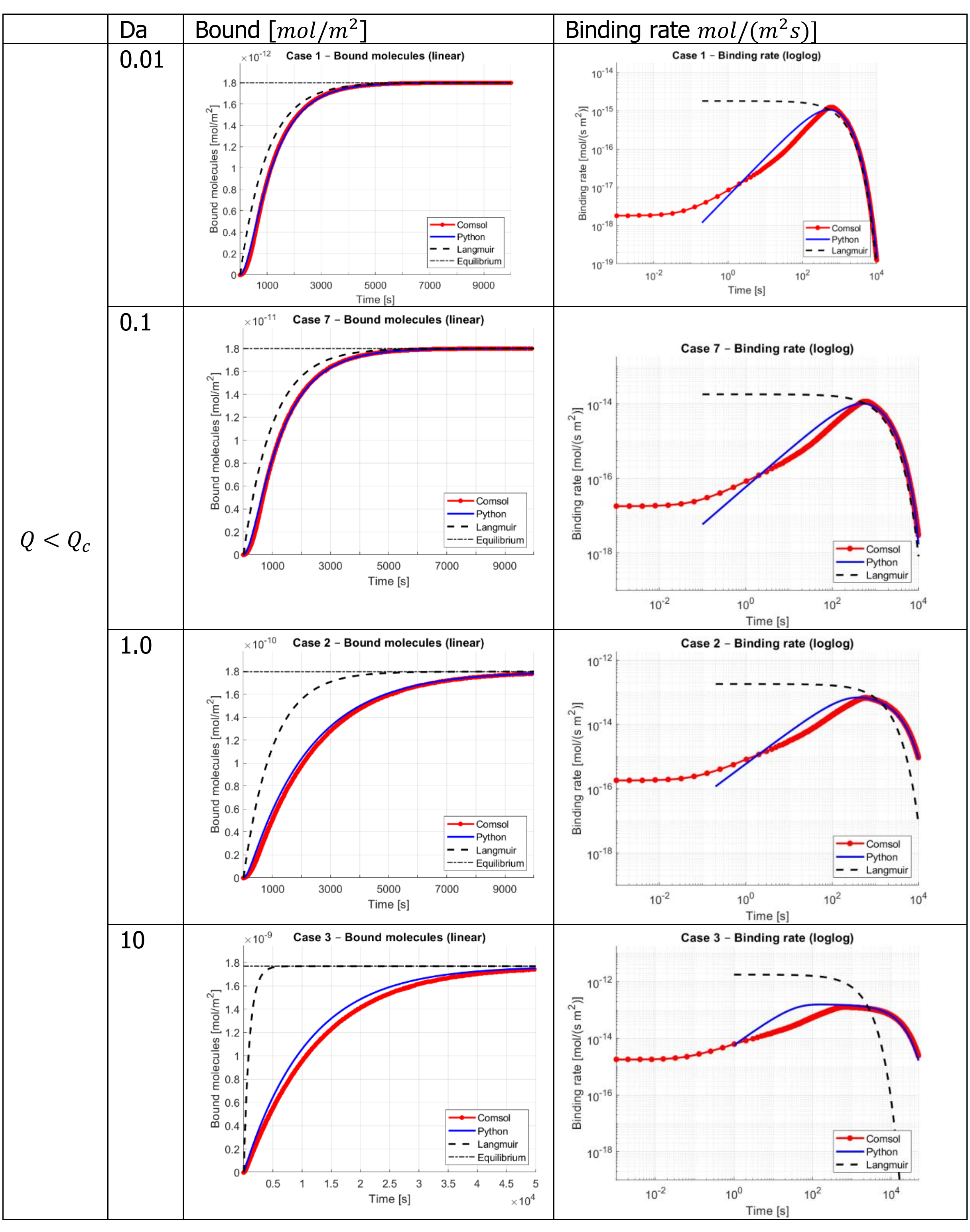

| | Da | Bound $[mol/m^2]$ | Binding rate $mol/(m^2s)]$ |
|---|---|---|---|
| $Q < Q_c$ | 0.01 | | |
| | 0.1 | | |
| | 1.0 | | |
| | 10 | | |

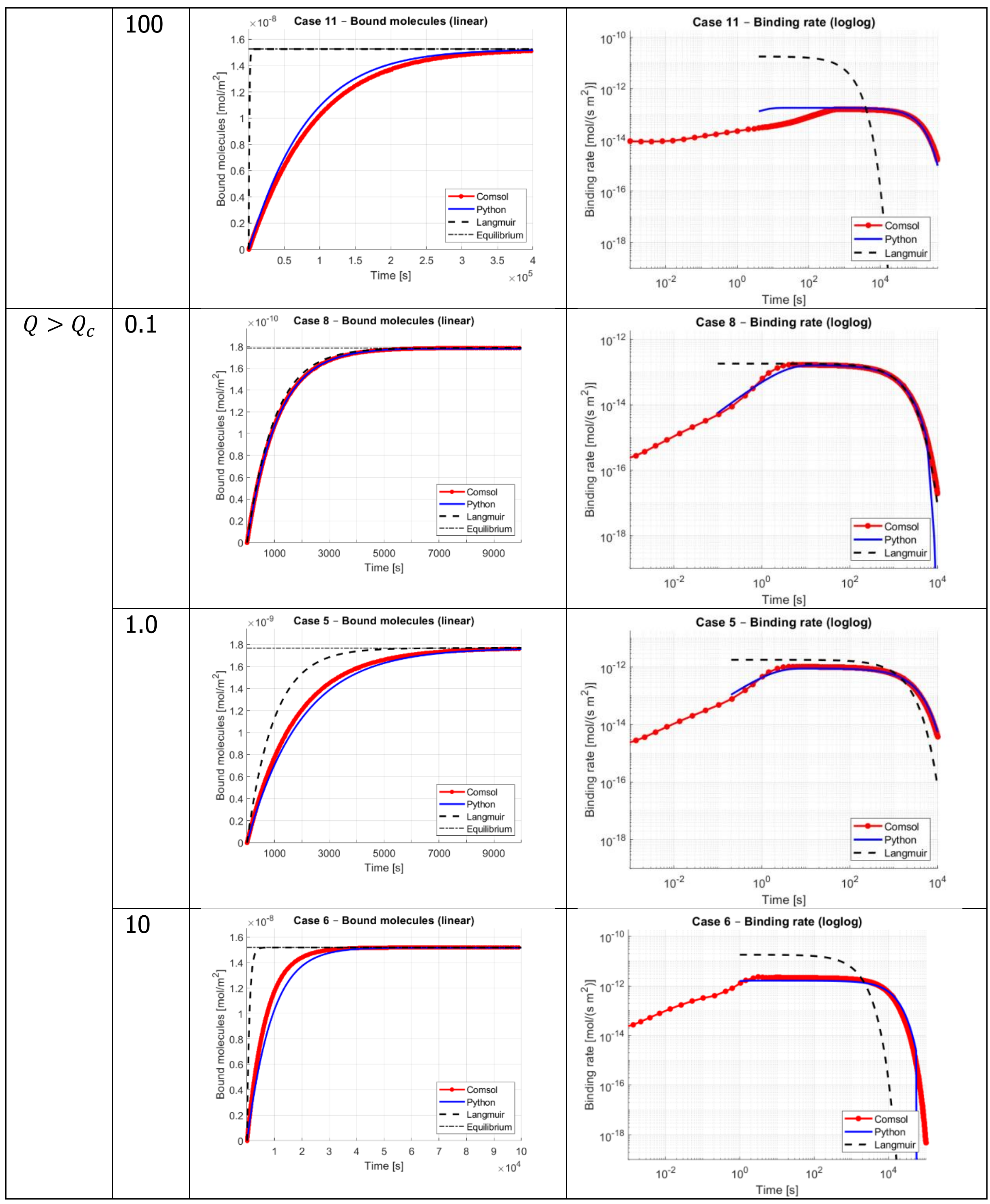

***Figure C.2*** *Kinetics comparison. For each of the simulated cases, the linear bound molecules and logarithmic binding rate are shown. Limited time stepping COMSOL results are shown in red, with our python model shown in blue. The (theoretical) Langmuir binding rate is shown in dotted black lines.*

The binding curves were validated by assessing the root-mean square error (RMSE) of our python simulations compared to the (limited time stepping) COMSOL simulations. The RMSE is calculated as shown before (eq. A1) and normalized to the characteristic values of $b_{eq}$ and $b_{eq}/t_R$ respectively:

$$\sigma_b = \frac{rmse_b}{b_{eq}} \cdot 100\% \quad \text{(C.1)}$$

$$\sigma_{dbdt} = \frac{rmse_{dbdt}}{b_{eq}/t_R} \cdot 100\% \quad \text{(C.2)}$$

The resulting errors $\sigma_b$ and $\sigma_{dbdt}$ are shown in figure C.3 for the chosen simulation points. The binding curves from our python model correspond well to COMSOL simulations for all regimes (reaction-limited, transport-limited, outside complete delivery). Indeed, the resulting error in $\sigma_b$ is less than 5% for all chosen simulations. The error is bigger for higher $Da$ values (transport-limited), due to the increasing importance of transport. For low values of $Da$ (reaction-limited), the biosensor follows Langmuir binding more closely and the binding error is smaller.

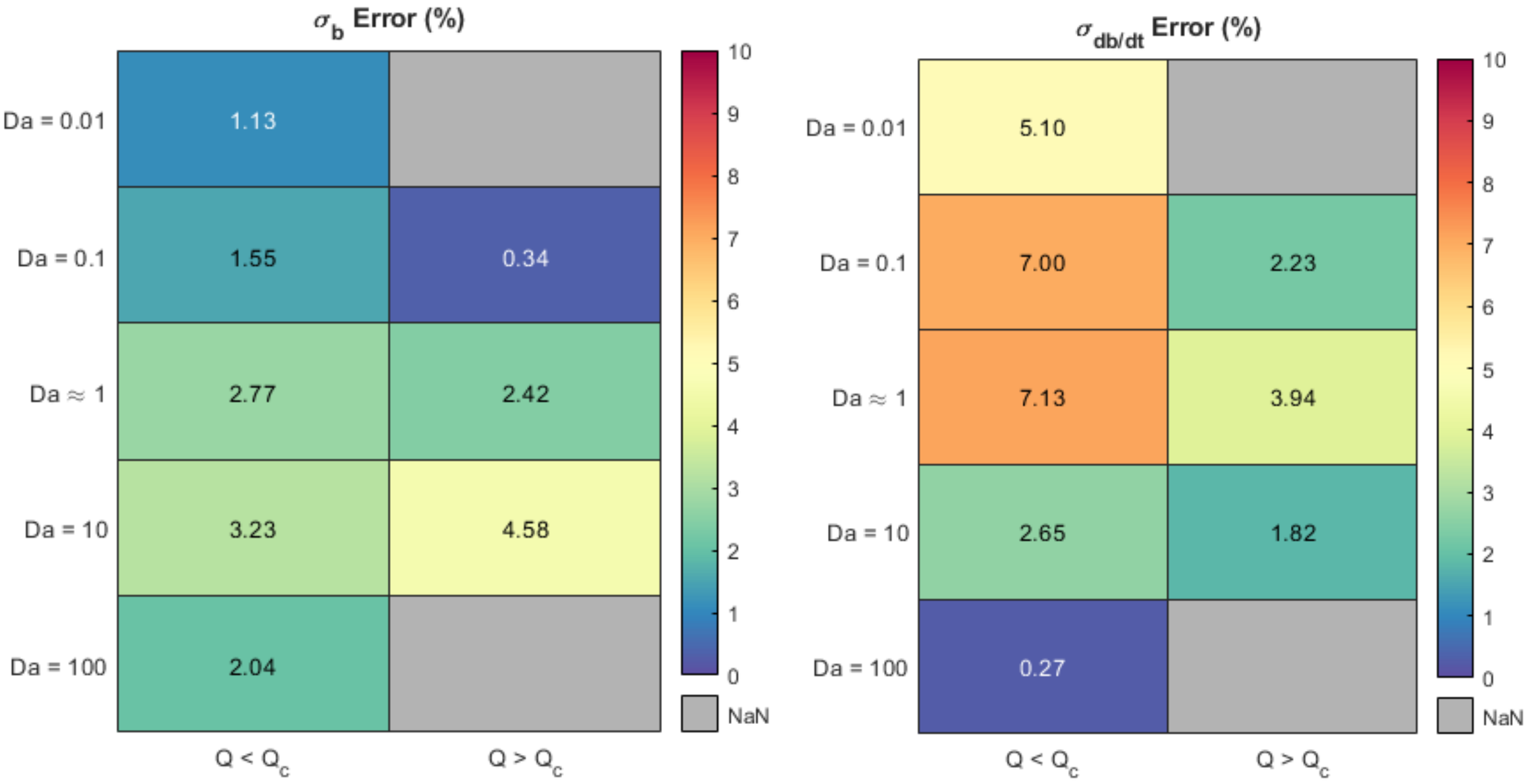

***Figure C.3*** *RMSE heatmaps for the bound molecule density (left) and binding rate (right).*

The biosensor equilibration time and required volume from simulations were compared to the analytical expressions. Tables C.5 and C.6 show the performance metrics for each of the simulated biosensors, as well as the characteristic $t_R$ and $V_{min}$. The analytical expressions (eq. 17 and eq. 21 from the main text) were used to obtain analytical values.

Each of the three models (analytical, python, COMSOL) were subsequently tested for agreement with the mean relative error (MRE):

$$MRE = \frac{1}{N} \sum_{i=0}^{N-1} \frac{|y_i - \hat{y}_i|}{|y_i|} \quad \text{(C.3)}$$

The resulting model validation curves are shown in figure C.4, with model results shown as points and a perfect agreement shown as a dotted line. The python model works well as a predictor for the performance metrics of $t_{eq}$ and $V_{req}$, with MRE values of 7.3% and 7.3% respectively. When using the analytical expressions as a predictor, the MRE values increase to 11.0% and 11.7% respectively.

| **ID** | $t_R$ | $t_{eq}$ **(analytical)** | $t_{eq}$ **(python)** | $t_{eq}$ **(comsol)** |
|---|---|---|---|---|
| 1 | 1000.0 | 3030 | 3392 | 3350 |
| 2 | 998.2 | 5950 | 6584 | 6700 |
| 3 | 982.3 | 32106 | 32620 | 36662 |
| 5 | 982.3 | 5885 | 5856 | 5340 |
| 6 | 848.17 | 27950 | 25354 | 20115 |
| 7 | 999.8 | 3294 | 3705 | 3679 |
| 8 | 998.2 | 3289 | 3290 | 3271 |
| 11 | 847.45 | 253880 | 228874 | 257605 |

***Table C.5*** *Equilibration time (seconds).*

| **ID** | $3V_{min}$ $[\mu L]$ | $V_{req}$ **(analytical)** | $V_{req}$ **(python)** | $V_{req}$ **(comsol)** |
|---|---|---|---|---|
| 1 | $5.39 \cdot 10^{-4}$ | 0.0550 | 0.0622 | 0.0614 |
| 2 | 0.0538 | 0.1082 | 0.1207 | 0.1228 |
| 3 | 0.5297 | 0.5831 | 0.5980 | 0.6721 |
| 5 | 0.5297 | 107.12 | 107.2 | 97.74 |
| 6 | 4.5482 | 500.30 | 453.84 | 360.07 |
| 7 | 0.0054 | 0.0593 | 0.0679 | 0.0674 |
| 8 | 0.0535 | 58.8806 | 58.906 | 58.557 |
| 11 | 4.5698 | 4.6159 | 4.1960 | 4.7228 |

***Table C.6*** *Required volume ($\mu$L).*

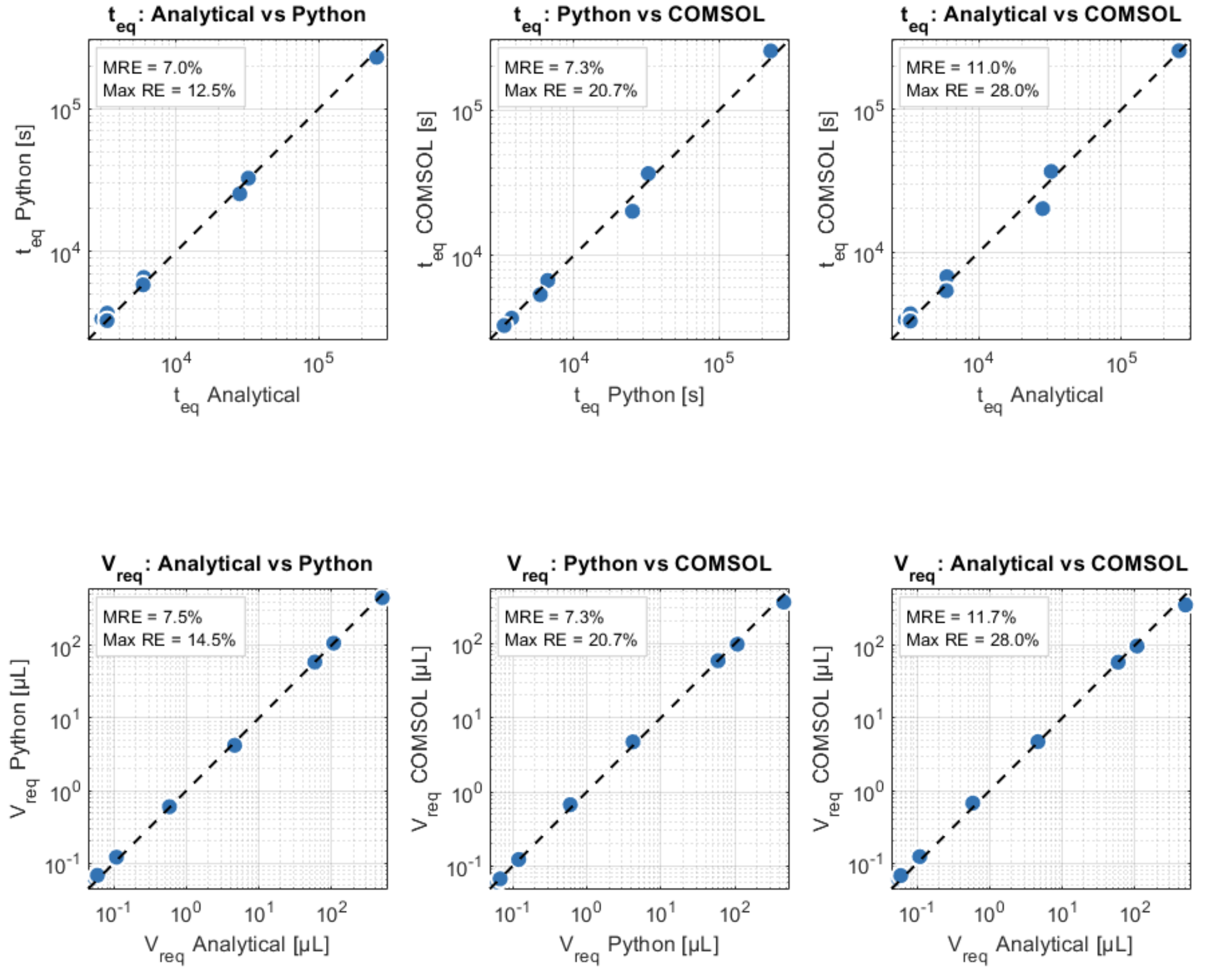

***Figure C.4*** *Model validation curves for performance metrics* $t_{eq}$ *and* $V_{req}$.

**Simulations Figure 5** (main text)

Simulations from our two-compartment model were performed on a range of biosensor geometries to obtain the points in figures 5a and 5b from the main text. The biosensor characteristics $(\lambda, c, Pe_H)$ used for these are shown in figures C.5 and C.6, obtaining 8 distinct systems. The bound probe density $b_m$ was varied to obtain a range of $Da$ for each of the systems.

The flow rate effects were simulated by fixing a system of $\lambda = 10$ and three values of $b_m$ (leading to $Da_c = 1, 10, 100$), and subsequently sweeping over increasing flow rates. From these sweeps (occurring outside complete delivery), the points outside complete delivery were obtained for figure 5b, as well as the points for figures 5c, 5d and 5e.

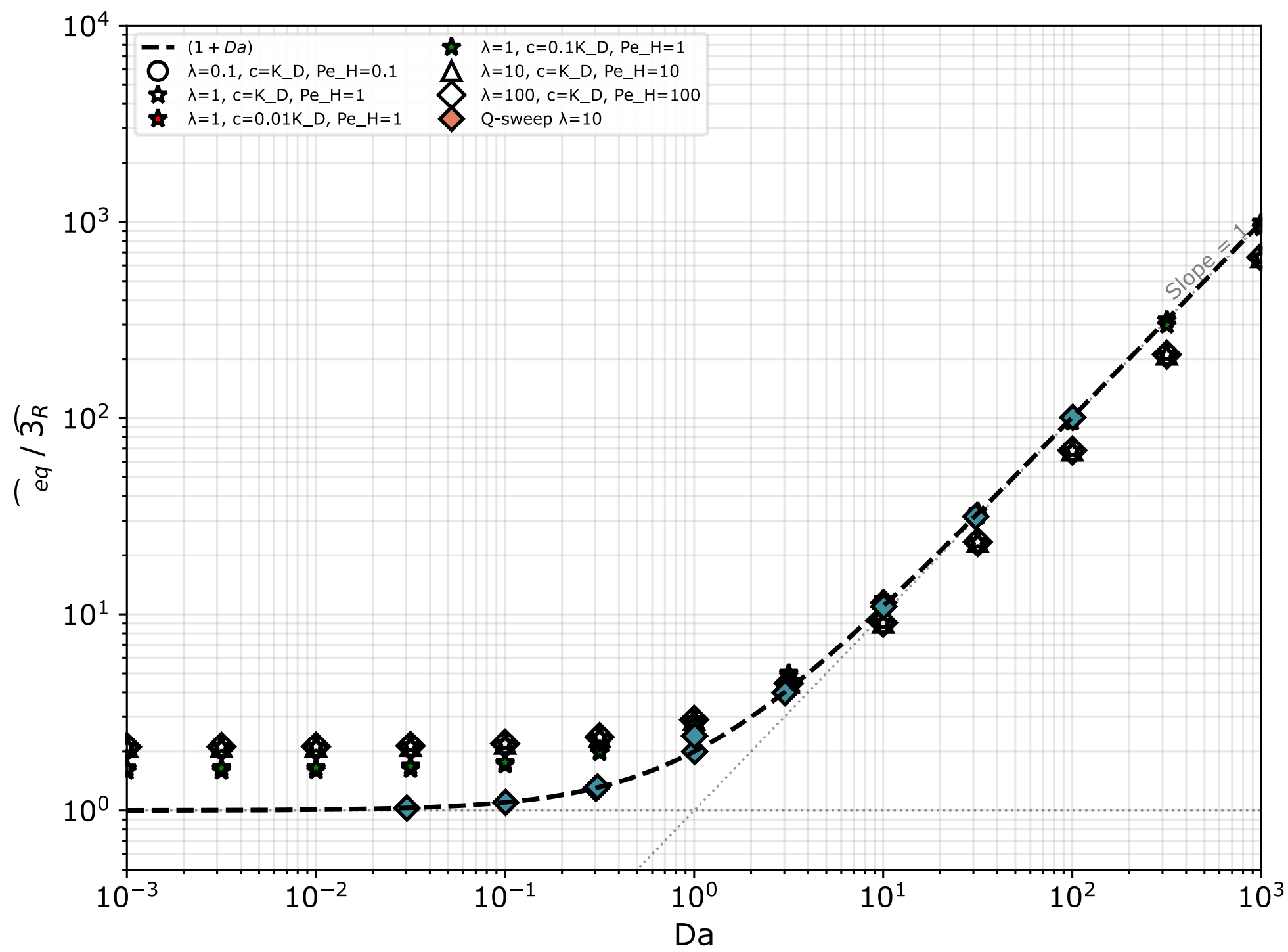

***Figure C.5*** *Two-compartment model simulations for equilibration time.*

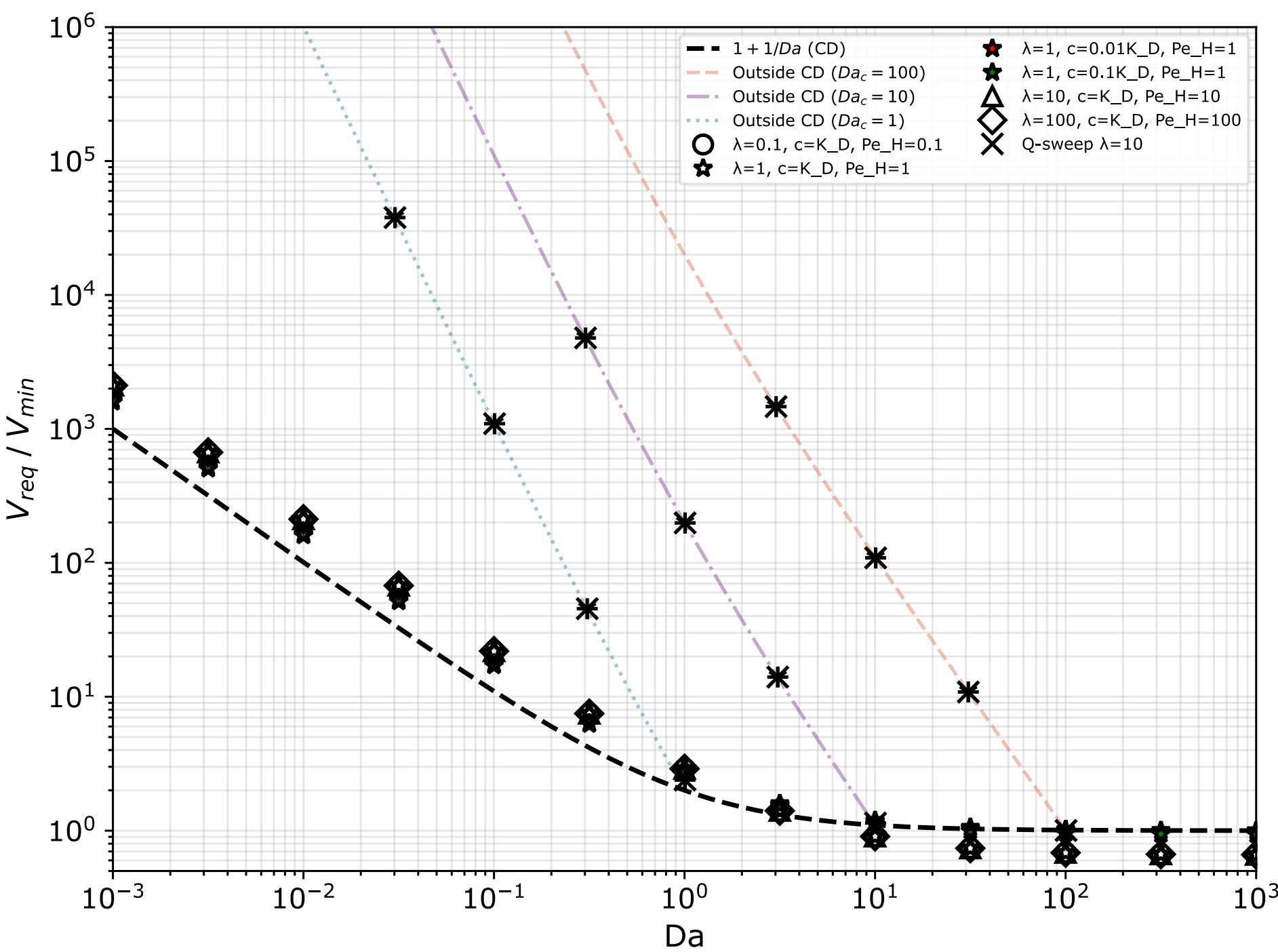

***Figure C.6*** *Two-compartment model simulations for required volume.*

# Supplementary D : Differential equations

Equations 11-13 in the main text describe the system of ODEs for the modeling of the biosensor. Below, the full description of the ODEs is given. The ODE for the interface concentration is described by:

$$V_s \frac{dc_s}{dt} = J_D - J_R - J_{c_s,out} \ (D.1)$$

$$V_s \frac{dc_s}{dt} = k_m W_s H_c (c_{in} - c_s) - \left(k_{on} c_s (b_m - b) - k_{off} b\right) W_s L_s \ \text{(D.2)}$$

The rate of binding is described by:

$$\frac{db}{dt} = k_{on} c_s (b_m - b) - k_{off} b \ \text{(D.3 )}$$

The rate of change in the non-interacting compartment is described by:

$$V_b \frac{dc_b}{dt} = J_{in} - J_D - J_{c,out} \ \text{(D.4 )}$$

$$\frac{dc_b}{dt} = \frac{1}{V} c_{in} Q - \frac{k_m}{L_s} c_{in} - \frac{1}{V} c_b \left(\frac{Q}{V} - \frac{k_m}{L_s}\right) \ \text{(D. 5)}$$

These systems of equations can then be simply solved by an ordinary differential equation (ODE) solver. For this model, it was additionally chosen to non-dimensionalize the equations to reduce small number effects. Each parameter was scaled to a respective starting value: $\hat{c} = c/c_{in}$ , $\hat{c_s} = c_s/c_{in}$, $\hat{b} = b/b_m$

The time scale is scaled to the residence time in the channel: $\tau = V/Q$, $\hat{t} = t/\tau$

Non-dimensionalization of Equations D.2, D.3 and D.5 leads to the system of equations:

$$\frac{d\hat{c}_s}{dt} = \frac{\tau k_m}{\delta} \frac{H_c}{L_s} (1 - \hat{c_s}) - \frac{\tau}{\delta} \left( k_{on} \hat{c}_s \left(1 - \hat{b}\right) - \frac{k_{off} b_m \hat{b}}{c_{in}} \right) (D.6)$$

$$\frac{d\hat{c}_b}{dt} = \frac{\tau}{V_b} \ (Q - k_m W_c H_c)(1 - \hat{c_b}) \ (D.7)$$

$$\frac{d\hat{b}}{d\hat{t}} = \tau\left(k_{on} c_{in} \hat{c}_s \left(1 - \hat{b}\right) - k_{off} \hat{b}\right) \ (D.8)$$

To simulate the biosensor, the system of ODEs from Equations D.6-8 is solved using pythons *solve_ivp*, with 'LSODA' solver. The error in the ODE system is described by the difference in molecules processed in the system compared to the number of molecules injected:

$$\zeta = \frac{N_{sys} - N_{inj}}{N_{inj}} \ (D.9)$$

# Supplementary E: Derivation of analytical expressions

### Surface concentration at the quasi-steady state

The ordinary differential equation for the concentration of the interface is described by (eq. D.2):

$$V_s \frac{dc_s}{dt} = k_m W_s H_c (c_{in} - c_s) - \left(k_{on} c_s (b_m - b) - k_{off} b\right) W_s L_s$$

In the quasi-steady state, the system reaches an equilibrium: $(dc_s)/dt = 0$. At the starting point when there is no bound molecules on the surface: b = 0:

$$k_m W_s H_c c_s + k_{on} c_s b_m W_s L_s = k_m W_s H_c c_{in}$$

$$c_s (k_m H_c + k_{on} b_m L_s) = k_m H_c c_{in}$$

$$c_s \left(1 + \frac{k_{on} b_m L_s}{k_m H_c}\right) = c_{in}$$

With $Da = k_{on} b_m L_s / k_m H_c$:

$$c_s (1 + Da) = c_{in}$$

$$c_s = c_{in} / (1 + Da)$$

### Equilibration time

The equilibration time of a biosensor is given by $t_{eq} \sim b_{eq}/(db/dt)$. With the binding rate $(db/dt)$ (eq. D.3) at initial time (b = 0):

$$\frac{db}{dt} = k_{on} c_s b_m$$

<u>For $Da \ll 1$</u>, $c_s \approx c_{in}$:

$$\frac{db}{dt} = k_{on} c_{in} b_m$$

With $b_{eq} = k_{on} c_{in} b_m / (k_{on} c_{in} + k_{off})$:

$$t_{eq} = -\ln(0.05) \frac{(k_{on}\, c_{in}\, b_m)/(k_{on}\, c_{in} + k_{off})}{k_{on} c_{in} b_m} \approx 3 t_R$$

<u>For $Da \gg 1$</u>:

$$c_s \approx \frac{c_{in}}{Da}$$

$$\frac{db}{dt} = \frac{k_{on} c_{in} b_m}{Da}$$

$$t_{eq} = -\ln(0.05) \frac{(k_{on}\, c_{in}\, b_m)/(k_{on}\, c_{in} + k_{off})}{k_{on} c_{in} b_m / Da} \approx 3 Da \cdot t_R$$

The factor $-\ln(0.05)$ is applied to obtain a 95% equilibrium, equivalent to the definition of the equilibration time as used in our two-compartment model (eq. 5 from main text).
First order interpolation of the found $t_{eq}$ values in reaction-limited and transport-limited regime obtains:

$$t_{eq} \approx 3(1 + Da)t_R$$

With the reaction timescale $t_R = \left(k_{on}c_{in} + k_{off}\right)^{-1}$.

**Required volume**

From Eq. 6, the required volume is defined as:

$$V_{req} = t_{eq}Q$$

Using the generalized expression for $t_{eq}$ from the previous section:

$$V_{req} = 3(1 + Da)t_R\, Q$$

<u>Inside complete delivery</u>, the Sherwood number is described by:

$$F = Pe_H$$

Leading to:

$$Q = k_m H_c W_c$$

Substituting $Q$ and $t_R$:

$$V_{req} = 3(1 + Da)(k_m H_c W_c)t_R$$

$$V_{req} = 3\left(\frac{k_m H_c W_c}{k_{on}c_{in} + k_{off}} + \frac{k_{on} b_m L_s W_c}{k_{on}c_{in} + k_{off}}\right)$$

With $W_s = W_c$ and

$$V_{min} = \frac{k_{on} b_m W_s L_s}{k_{on}c_{in} + k_{off}}$$

We obtain:

$$V_{req} = 3(V_{min} + \frac{1}{Da})$$

Leading to:

$$\frac{V_{req}}{3V_{min}} = \frac{1}{Da} + 1$$

Outside complete delivery:

$$Pe_s = 6\lambda^2 Pe_H$$

The critical $Pe_{H,c}$ is found at the intersection o

$$F = Pe_H$$
$$Pe_{H,c} = 0.81\, Pe_{s,c}^{1/3} = 0.81\left(6\lambda^2 Pe_{H,c}\right)^{1/3}$$

$$Pe_{H,c} = \left(0.81(6\lambda^2)^{1/3}\right)^{3/2} \approx 1.79\lambda$$

Damkohler number for crossover point $Q = Q_c$:

$$Da_c = \frac{k_{on} b_m L_s}{k_{m,c} H_c} = \frac{k_{on} b_m L_s H_c}{F_c D H_c} = \frac{k_{on} b_m H_c}{1.79\, D}$$

$$F \approx 0.81 Pe_s^{1/3} = 0.81(6\lambda^2)^{1/3} \cdot Pe_H^{1/3} = Pe_{H,c}^{2/3} \cdot Pe_H^{1/3}$$

With $Pe_H = \frac{Q}{W_c D}$,

$$Q_c = Pe_{H,c} W_c D \approx 1.79\lambda W_c D$$

The effect of flow rate on Da can be found by comparing the ratio of $Da/Da_c$:

$$\frac{Da}{Da_c} = \frac{k_{m,c}}{k_m} = \frac{F_c}{F} = \frac{Pe_{H,c}}{Pe_{H,c}^{2/3} \cdot Pe_H^{1/3}} = \left(\frac{Pe_{H,c}}{Pe_H}\right)^{1/3} = \left(\frac{Q_c}{Q}\right)^{1/3}$$

$$Da = Da_c \left(\frac{Q_c}{Q}\right)^{1/3}$$

Rewrite as:

$$Da^3 = Da_c^3 \left(\frac{Q_c}{Q}\right)$$

$$Q = Q_c \frac{Da_c^3}{Da^3}$$

Substitute into $V_{req}$:

$$V_{req} = t_{eq} Q = (1 + Da)\tau_R \cdot Q_c \cdot \frac{Da_c^3}{Da^3}$$

Using:

$$Q_c = Pe_{H,c} W_s D = F_c W_s D = \frac{k_{on} b_m L_s W_s}{Da_c}$$

We obtain:

$$V_{req} = 3(1 + Da)t_R \frac{k_{on}b_m L_s W_s}{Da_c} \frac{Da_c^3}{Da^3}$$

$$V_{req} = 3(1 + Da)V_{min} \frac{Da_c^2}{Da^3}$$

Leads to:

$$\frac{V_{req}}{3V_{min}} = \left(\frac{1}{Da} + 1\right) \frac{Da_c^2}{Da^2}$$